\documentclass{aa}
\usepackage{graphics,psfig,times}
\begin{document}

\title{The evolution of interstellar clouds in a streaming hot
       plasma including heat conduction}


\author{W. Vieser \inst{1,2}
\and
        G. Hensler \inst{1}
}

\offprints{G. Hensler,\\ \email{hensler@astro.univie.ac.at}}

\institute{Institute of Astronomy, University of Vienna, T\"urkenschanzstr.\ 17,
           A--1180 Vienna, Austria 
\and
           Christoph-Probst-Gymnasium, Talhofstr.\ 7, D--82205 Gilching, Germany
}

\date{Received 30 January 2005/ Accepted 20 October 2006}

 
  \abstract
{The interstellar medium contains warm clouds that are embedded in a hot
dilute gas produced by supernovae. Because both gas phases are in
contact, an interface forms where mass and energy are exchanged.
Whether heat conduction leads to evaporation of these clouds or 
whether condensation dominates has been analytically derived. 
Both phases behave differently dynamically 
so that their relative motion has to be taken into account. }
{Real clouds in static conditions that experience saturated heat 
conduction are stabilized against evaporation if self-gravity and cooling 
play a role. Here, we investigte to what extent heat conduction can 
hamper the dynamical disruption of clouds embedded in a streaming hot plasma. }
{To examine the evolution of giant molecular clouds in the stream 
of a hot plasma we performed two-dimensional hydrodynamical simulations 
that take full account of self-gravity, heating and cooling effects and 
heat conduction by electrons.
We use the thermal 
conductivity of a fully ionized hydrogen plasma proposed by Spitzer
and a saturated heat flux according to Cowie \& McKee in regions where the
mean free path of the electrons is large compared to the temperature
scaleheight. }
{Significant structural and evolutionary differences occur between simulations
with and without heat conduction. Dense clouds in pure dynamical models 
experience dynamical destruction by Kelvin-Helmholtz (KH) instability. 
In static models heat conduction leads to evaporation of such clouds. 
Heat conduction acting on clouds in a gas stream smooths out steep 
temperature and density gradients at the edge of the cloud because the
conduction timescale is shorter than the cooling timescale. This
diminishes the velocity gradient between the streaming plasma and the
cloud, so that the timescale for the onset of 
KH instabilities increases, and the surface of the cloud becomes less
susceptible to KH instabilities. 
The stabilisation effect of heat conduction against KH instability is more
pronounced for smaller and less massive clouds.
As in the static case more realistic cloud conditions allow heat
conduction to transfer hot material onto the cloud's surface and
to mix the accreted gas deeper into the cloud. }
{In contrast to pure dynamical models of clouds in a plasma and to analytical 
considerations of heat conduction that can evaporate such clouds embedded in a hot plasma, 
our realistic numerical simulations demonstrate that this destructive effect 
of KH instability is significantly slowed by heat conduction so that clouds 
can survive their passage through hot gas.}


\titlerunning{Evolution of streaming IS clouds with heat conduction}

\maketitle


\section{Introduction}

\subsection{The multi-phase Interstellar Medium}

The Interstellar Medium (ISM) is frequently described as an inhomogeneous
ensemble of three phases (Mc~Kee \& Ostriker 1977: \cite{mo77}).
The cold neutral phase with temperature $T \sim 80 \mbox{ K}$ and
density $n \sim 40 \mbox{ cm}^{-3}$ is represented by the cores of
molecular clouds which are confined by a warm neutral to slightly ionized
medium ($T \sim 8000 \mbox{ K}$, $n \sim 0.3 \mbox{ cm}^{-3}$).
These two components
are in pressure equilibrium if the gas is externally heated and can cool
radiatively (Field et al. \cite{f69}). According to \cite{mo77} they are
embedded in a third phase of the ISM:
the hot dilute intercloud medium (HIM) with $T \sim 10^6 \mbox{ K}$ and
$n \sim 10^{-3} - 10^{-4} \mbox{ cm}^{-3}$. Since this
component is produced locally by supernova explosions at even higher
temperatures and densities, originally it cannot
be in pressure equilibrium with the cooler phases and therefore has to
expand. During this expansion
shocks arise and the HIM penetrates the ambient clumpy
ISM. Denser clouds cannot be swept-up by the shock front but are overrun
and become engulfed by the HIM. 
Because of hot gas expansion the relative velocity between
both gas phases leads to strong distortion of the clouds, resulting in
a stripped-off gas tail. Clouds of this type can be found as remnants 
of larger clouds in galactic chimneys such as the one associated with the
H{\sc ii} region W4 (\cite{he96}; \cite{ta99}) or behind shock fronts 
of supernovae. 
Due to the strong discrepancy in the
physical states between the phases an interface has to form where the
hot phase and the molecular cloud are in contact. Temperatures and densities
are connected through steep gradients that lead to energy and mass transfer.

This situation occurs in many
astrophysical phenomena such as {\it{High-Velocity Clouds}} 
(HVCs), {H{\sc i}} structures characterized by radial velocities that are
incompatible with simple models of the differential rotation of the galactic
disk (see Wakker \& van Woerden (\cite{ww97}) for a recent review). 
Interferometer
measurements of the 21 cm {H{\sc i}} line at 1' resolution show several
small clumps embedded in larger emission regions 
(Wakker \& Schwarz \cite{ws91}). This, along with
their line width, led Wakker \& Schwarz to conclude that the HVCs have
a multi-phase structure consisting of a cold, dense core and a warmer, more
tenuous halo. Distance measurements of the cloud complexes remain difficult.
For at least two of them upper limits for their distances are deduced by
Danly et al. (\cite{d93}), Keenan et al. (\cite{k95}) and
van Woerden et al. (\cite{w97}) which place them
in the galactic halo. From absorption line and X-ray observations
it is well established that the galactic halo is filled with hot gas
with a scaleheight of 4 kpc (\cite{pi98})
so that the HVCs have to interact with the hot gas.
{\it{ROSAT}} observations show an increase of X-ray emission at the
edges of the clouds that can be assumed as evidence for the existence of a
hot interface (Kerp et al. \cite{k94}, \cite{k295}).
The detection of {\it{velocity bridges}} (\cite{pi96}),
connections of {H{\sc i}} gas in velocity-space-diagrams, could also stem
from the interaction of HVCs with low or intermediate-velocity gas.
Although the origin of most of the complexes is still speculative, HVCs are
a classical example of cold multi-phase structures
moving through a hot plasma.

Another possibility for the scenario described above ismuch more massive and 
larger proto-globular cluster clouds (PCCs) with temperatures near $10^4$ K and
densities several hundred times that of the surrounding gas. They therefore
are gravitationally unstable at Jeans masses larger than
$10^6$ M$_{\sun}$. An upper limit for their mass can be approximated by
the critical mass for an isothermal sphere
that is embedded in a surrounding medium with
pressure $P_{\mbox{\tiny{ISM}}}$ (Ebert \cite{e55}; Bonner \cite{b56}):

      \begin{equation}{\label{mmax}}
      M_{\mbox{\tiny max}} = 1.2
      \left (
      \frac{k_BT}{\mu}
      \right )^2
      G^{-3/2} P^{-1/2}_{\mbox{\tiny{ISM}}}
      \end{equation}

where $G$ is the gravitational constant, $k_B$ Boltzmann's constant,
$T$ a PCC temperature and $\mu$ its mean molecular weight.
These clouds are assumed to originate from condensations of
thermally unstable gas with temperatures of some million Kel\-vin in the
early epoch of galaxy formation and can be envisaged
as progenitors of globular clusters (Fall \& Rees \cite{fr85}). 
The PCCs had to resist their gravitational collapse 
until star formation ignites with a very high efficiency.
This condition is not easily fulfilled because the PCC is
accelerated by the gravitational potential of the forming protogalaxy. While
moving through the hot halo it will become subject to the
growth of Kelvin-Helmholtz (KH) and Rayleigh-Taylor (RT) instabilities
(Drazin \& Reid \cite{dr81}).

      \subsection{Numerical models of clouds in a hot plasma stream}

To investigate an influence of KH and RT instabilities on the
evolution of molecular clouds several authors have used numerical
simulations. Murray et al. (\cite{m93}) compared the evolution of clouds embedded
in the subsonic stream of a dilute medium with and without self-gravity.
Radiative losses are neglected and the gravitational potential of the
clouds remain fixed to the initial value of the calculation.
Models without self-gravity and cloud masses
$M_{\mbox{\tiny cl}} \ll M_{\mbox{\tiny max}}$ break up
after only a few dynamical times $\tau_{\mbox{\tiny dyn}} =
R_{\mbox{\tiny cl}}/c_{\mbox{\tiny cl}}$, where $R_{\mbox{\tiny cl}}$ is the cloud
radius and $c_{\mbox{\tiny cl}}$ is the sound speed inside the cloud. The mass loss
after 2.5 $\tau_{\mbox{\tiny dyn}}$ is 20\% and increases to
75\% after 3.8 $\tau_{\mbox{\tiny dyn}}$. On the
other hand, gravitationally bound clouds with a gravitational
acceleration $g = g_{\mbox{\tiny crit}}$,
the value where perturbations with
wavelengths of the order of $R_{\mbox{\tiny cl}}$ are totally damped,
show a different evolution. 
The value for $g_{\mbox{\tiny crit}}$ results from the relative
velocity $U$, the cloud density $\rho_{\mbox{\tiny cl}}$ and the
density of the background material $\rho_{\mbox{\tiny bg}}$ to
$g_{\mbox{\tiny crit}} = 2 \pi U^2 \rho_{\mbox{\tiny bg}} / (
\rho_{\mbox{\tiny cl}} R_{\mbox{\tiny cl}} ) $.

The gravity is sufficient to stabilize the cloud
although the cloud edge is distorted. After 3.2 $\tau_{\mbox{\tiny dyn}}$
only 2\% of
the initial cloud mass is lost and after 10 $\tau_{\mbox{\tiny dyn}}$, 11\%.
Severing (\cite{s95}) improved the simulations done by Murray et al.
(\cite{m93}) by solving the Poisson equation for self-gravity self-consistently
at each timestep and by
including heating and cooling effects. The cooling takes a collisionally
dominated plasma in thermal and ionization equilibrium into account
(B\"ohringer \& Hensler \cite{bh89}) and is balanced by heating for a cloud at rest. He demonstrated that clouds in a stream with Mach number $0.1$ and
with $g > g_{\mbox{\tiny crit}}$ gain mass while at higher but still
subsonic Mach numbers and with $g < g_{\mbox{\tiny crit}}$ similar results
to those by Murray et al. are achieved.
Dinge (\cite{d97}) applied cooling only to the gas ablated from the cloud and
showed that
cooling tends to accelerate the destruction of the cloud by RT instability.
However, this arises due to the fact that the stripped cloudlets hit
the cloud from the rear and thus trigger soundwaves which move
through the cloud
and stretch it along the symmetry axis. Globally, his results agree
qualitatively with those of the former authors.

Vietri et al. (\cite{v97}) examined the influence of radiative losses on the
evolution of KH instabilities. By analytic approximations they proved that
plane parallel identical fluids are destabilized if they stream
relative to each other at
high Mach numbers while low Mach number flows tend to be
more stable. For fluids having different densities, cooling processes
exacerbate the KH instabilities and all Mach numbers become unstable
although with moderate growth rates. For clouds without
self-gravity this means that KH instabilities cannot be suppressed.
The instability is constrained to a small volume around the
surface because cooling and heating timescales are shorter than the dynamical
ones so that the instability cannot extend much beyond the interface.

In addition, observationally derived structure parameters allow the 
conclusion that HVCs are bound by a significant Dark Matter halo mass. 
\cite{qm01} studied with hydrodynamical simulations the behavior 
of HVCs passing the dilute hot halo gas of a typical disk galaxy at large 
distances. The main purpose of their numerical models was to compare the  
head-tail structure with observations to study the 
necessity for a Dark Matter content and a possible lower density limit 
of the surrounding intergalactic medium (IGM). 
Although they do not trace the evolution with respect 
to mass loss and structure survival (e.g. the pure gas clouds are shown 
after 3 Myr only), the main conclusions are two-fold, namely, at first 
that the observed head-tail structures with sufficient densities are 
produced if the wind density of the IGM exceeds 
n$_{IGM} \geq 10^{-4}$ cm$^{-3}$ due to ram-pressure stripping (RPS) 
but independent of the existence of a DM halo. 
This supports the galactic fountain scenario for their origin 
because the outermost IGM is expected to fall short of this density threshold. 
Secondly, the velocity was varied between 200 and 400 km s$^{-1}$ without 
any substantial change in the structure.

      \subsection{Heat conduction}

Stable models for Giant Molecular Clouds (GMCs) like HVCs and PCCs
assuming hydrostatic and thermal equilibrium
consist of large temperature and density gradients at the
surfaces of the clouds where the energy densities of the surrounding ISM
and the cloud become equal. 

There are indications that the
observed {H{\sc i}} masses of some HVCs are orders of magnitude too low
to provide gravitational stabilization, so they must be confined by
external pressure (Konz et al. \cite{k02}). On the other hand, the core-halo
structure seen in some HVCs (e.g Wolfire et al. \cite{w95}) may
result from minihalos composed of gas and dark matter that would
likely move together as the clouds fall onto the Milky Way 
(Blitz et al. \cite{b99}). In this picture, the largest HVCs are 
gravitationally bound by the dark matter rather than pressure 
confined (\cite{bb99}, 2000). 

Because of the large temperature gradients and the high temperature of
the surrounding medium of the order of some million Kelvin, heat conduction
must play a substantial r\^ole in the evolution of such clouds.

Analytical studies of the influence of heat conduction on the evolution
of clouds at rest were undertaken by several authors:
Cowie \& McKee (1977) (hereafter: \cite{cm77})
investigated the evaporation of spherical clouds in general. For the classical
regime described by a collision-dominated plasma they used the conductivity
of Spitzer (1962). In this case,
the heat flux due to heat conduction can be written as

      \begin{equation}
      \vec{q}_{\mbox{\tiny class}} = - \kappa \cdot \vec{\nabla} T
      \end{equation}

with

      \begin{equation}
      \kappa = \frac{1.84 \cdot 10^{-5} T^{5/2}}{\ln \Psi}
      \mbox{erg s$^{-1}$ K$^{-1}$ cm$^{-1}$} \quad ,
      \end{equation}

      where the Coulomb logarithm is

      \begin{equation}
      \ln \Psi = 29.7 + \ln \left [ \frac{T_{6,e}}
      {\sqrt{n_e} } \right ] \quad ,
      \end{equation}

with the electron density $n_e$ (in cm$^{-3}$) and the electron 
temperature $T_{6,e}$ in units of $10^6$ K.

In a collisionless plasma the diffusion approximation for the heat flux
breaks down because the mean free path of electrons becomes comparable 
to or even larger than the temperature scaleheight. 
In this so-called saturated regime described by a collisionless
plasma, CM77 used a flux-limited heat flux.
This takes charge conservation into account
and yields results in good agreement with more sophisticated treatments
(e.g. Max et al. \cite{mmm80}) and with numerical simulations of laser
heated plasmas (Morse \& Nielsen \cite{mn73}; Mann\-heimer \&
Klein \cite{mk75}).
The saturated heat flux as an upper limit takes the form
(\cite{cm77})

      \begin{equation}{\label{qsat}}
      |\vec{q}_{\mbox{\tiny sat}}| = 5 \Phi_s \rho c^3
      \end{equation}

with the sound speed $c$. $\Phi_s$ is an efficiency factor less
than or of the
order of unity, which embodies some
uncertainties connected with the flux-limited treatment and 
flux suppression by magnetic fields ($\Phi_s =1$ in our
calculations).

For an abrupt change of the conductivity from the classical
to the saturated regime this leads to an envelope around
the cloud consisting of three layers: a saturated zone embedded in an
inner and outer classical
zone. CM77 obtain a classical mass-loss rate of

      \begin{equation}
      \dot{m}_{\mbox{\tiny class}} = \frac{16 \pi \mu \kappa_f R}{25 k_B}
      \end{equation}

where $\kappa_f$ is the conductivity evaluated for the unperturbed hot
medium far away from the cloud. $R$ is the cloud radius and $\mu$ the
mean molecular weight. The mass-loss rate is
lower for the saturated case. In McKee \& Cowie (\cite{mc77}) they included
radiative losses in their studies.
As a criterion to separate the classical and the saturated case for a cloud
of radius $R$ embedded in a hot gas with temperature $T_f$ and electron
density $n_{ef}$, \cite{cm77} introduced a global saturation
parameter $\sigma_0$ which is the ratio of the electron
mean free path to the cloud radius

      \begin{equation}{\label{sigma}}
      \sigma_0=\frac{0.08 \kappa_f T_f}{\Phi_s \rho_f c^3_f R}
      = \frac{1.23 \cdot 10^4 T^2_f}{n_{ef} R} \qquad .
      \end{equation}

For $\sigma_0 < 0.027/\Phi_s$ material condenses onto the cloud because
radiative losses exceed the conductive heat input.
For $0.027/\Phi_s < \sigma_0 \le 1$ the cloud suffers
classical evaporation, while for $\sigma_0 > 1$ the evaporation is
saturated. McKee \& Begelman (\cite{mb90}) found similar results
introducing the
so-called Field length

      \begin{equation}{\label{field}}
      \lambda_{\mbox{\tiny F}} = \left[
      \frac{ \kappa T}{n^2 {\cal{L}}_M}
      \right]^{1/2}
      \mbox{with} \quad
      {\cal{L}}_M \equiv \max(\Lambda,\Gamma/n) \quad ,
      \end{equation}

in which the cooling or heating is
comparable to the conductive energy exchange.
For $R > \lambda_{\mbox{\tiny F}}$ condensation occurs, otherwise
the cloud will be evaporated. $n$ is the particle density,
$\Lambda$ the cooling rate, $\Gamma$ the heating rate.

These studies concluded that cold clouds in a dilute medium at a temperature
of some million Kelvin always experience evaporation,
but none of them examined
more realistic multi-phase clouds or included self-gravity.

In a seperate paper (Vieser \& Hensler 2005; hereafter: \cite{vh05}) we
studied the differences caused by heat conduction on the
eva\-pora\-tion/con\-den\-sation competition between
the fixed analytical description and the more realistic approach of flux
saturation that adopts flexibly to the temporal physical state.

The main results of these investigations are as follows: The
analytical mass loss rates of a cloud at rest in a hot and rarefied medium 
can be reproduced in numerical simulations for the pure classical case, 
because the evaporated material is pushed away with supersonic speed. 
The initial large density and temperature jump
at the edge of the cloud remains unaltered during the calculation.
Taking the more realistic saturated heat
flux into account, a transition zone forms at the cloud edge in which
the steep temperature and density gradients are reduced. 
This results in a lower evaporation rate than predicted. 
Simulations that include additional heating and cooling show an even more
dramatic effect. The clouds can even gain material if radiative cooling 
exeeds the energy input by heat conduction.   

Here we examine the evolution of molecular clouds in
the stream
of a hot, dilute medium. 
The treatment of heat conduction in the context of hydrodynamical simulations
is described in \S2. Analytical estimates of the influence of heat
conduction are compared with the results of
dynamical models with and without heat conduction in \S3.
Conclusions are drawn in \S4.

      \section{Hydrodynamical treatment}

The evolution of clouds in the subsonic stream of a hot plasma is studied
by two-dimensional hydrodynamic simulations. 

The hydro-part of this Eulerian, explicit code is based on the 
prescription of Rozyczka (1985) and has been extensively tested and used by
different authors (e.g. Yorke \& Welz \cite{yw96}). The hydrodynamic
equations have been formulated in cylindrical coordinates (r, z), 
assuming axial symmetry around the z-axis that is also the flow
direction. The cloud's center is located on the z-axis.
The differencing scheme used to discreticize the equations is 
second-order accurate in space because a ``staggered grid'' is used . 
We applied operator splitting for time integration, because 
numerical experiments have shown that a multi-step solution procedure 
is more accurate than a single integration step based on preceding values
(\cite{sn92}). 
The advection scheme of van Leer (\cite{vl77}) 
is employed. Since the basic code is explicit, the 
Courant-Friedrichs-Lewy (CFL) condition determines the maximum
time step for the hydro-part. Because the conduction time step is
smaller than the CFL one, the temperature distribution has to be 
calculated several times in one hydro time
step. Von Neumann-Richtmyer artificial viscosity is used for the
treatment of shocks. In order to prevent the cloud from moving out
of the computational domain due to drag forces, the cloud center of
mass is re-adjusted at each time step.

The grid parameters, the resulting physical domain
and the resolution are listed in Table~\ref{t1} for three representative
models. The grid resolution is 28 - 33 zones per initial cloud radius.

      \begin{table}[ht]
      \centering
      \caption{\label{t1}Numerical parameters for the simulations}
      \begin{tabular}{llll}
      \hline
      & numerical grid
      & physical grid
      & resolution \\
      Model
      & cell$^2$
      & pc$^2$
      & pc cell$^{-1}$ \\ \hline \hline
      U & $640 \times 200$ & $800 \times 250$ & 1.25 \\
      K & $640 \times 200$ & $80 \times 25$ & 0.125 \\
      E & $640 \times 200$ & $50 \times 15.625$ & $7.8 \cdot 10^{-2}$ \\
      \end{tabular}
      \end{table}

The boundary conditions on the upper and right-hand sides are semi-permeable 
to allow for an outflow of gas from the computational domain.
The physical parameters at the lower boundary, the symmetry
axis, are mirrored. The parameters, density and temperature at the left-hand
boundary are fixed and the inflow of the plasma is initialized at a constant
value. In order to trace the condensation of the streaming
material onto the cloud a new quantity, ``colour'', is introduced
that is set in each cell to the density fraction of hot
ISM. At the beginning only the cells around the cloud possess a
non-zero ``colour'' of value unity. During the
calculation this quantity is advected like the others, such as
mass density or energy density.
The Poisson equation for self-gravity
was solved at each tenth timestep because significant density changes
of the cloud
structure happen on a much longer timescale than the dynamical one.
The energy equation includes
heating, cooling and heat conduction:

      \begin{equation}
      \frac{\partial e}{\partial t} +
      \vec{\nabla} \cdot (e \vec{v}) =
      -P \vec{\nabla} \cdot \vec{v} + \Gamma - \Lambda - \vec{\nabla} \cdot
      \vec{q} \quad .
      \end{equation}

Here $e$ denotes the energy density, $\vec{v}$ the velocity, $P$ the pressure,
$\Gamma$ the heating rate, $\Lambda$ the cooling rate and $\vec{q}$ the
heat flux. The equation of state for an ideal gas is assumed to be valid:

      \begin{equation}
      P=(\gamma - 1)e \qquad \mbox{with} \qquad \gamma = 5/3
      \end{equation}

The cooling function used assumes collisional ionisation equilibrium and
is a combination of the function introduced by B\"ohringer \& Hensler (\cite{bh89})
for $T>10^4$K and solar metallicity and by
Dalgarno \& McCray (\cite{dm72}) for the lower
temperature regime.
The heating function considers cosmic rays (Black \cite{b87}),
X-rays and the photoelectric effect on dust grains (de Jong \cite{dj77}
; de Jong et al. \cite{dj80}).

The heat flux is calculated by taking both the classical and the
saturated flux into account. In order to apply a smooth
transition between the classical and saturated regimes we use the
analytical form by Slavin \& Cox (\cite{sc92})

      \begin{equation}
      \vec{q}=|\vec{q}_{\mbox{\tiny sat}}| \left(
      1- \exp \left[ -\frac{\vec{q}_{\mbox{\tiny class}}}
      {|\vec{q}_{\mbox{\tiny sat}}|}
      \right]
      \right) \; .
      \end{equation}

This guarantees that the lower flux is taken if both differ significantly.
The heat flux due to electron diffusion is calculated separately using an
implicit method which follows a scheme introduced by
Crank \& Nicolson (\cite{cn47})
and Juncosa \& Young (\cite{jy71}). To couple the two directions in space the
method of fractional steps by Yanenko (\cite{y71}) is used.

A detailed description of the implementation of heat conduction in
the existing hydro-code is given in \cite{vh05}, as well as
a comparison of analytical solutions and numerical test cases to prove
the reliability of the code.

The initial temperature and density profiles of the clouds are generated
for hydrostatic and thermal equilibrium under the constraint of spherical
symmetry:

      \begin{equation}{\label{gl10}}
      \vec{\nabla}\Phi = - \frac{1}{\rho} \,\vec{\nabla} P
      \end{equation}

      \begin{equation}{\label{gl11}}
      \Gamma(\vec{r}) = \Lambda(\vec{r}) \; ;
      \end{equation}

$\Phi(r)$ is the gravitational potential.
Setting the temperature $T_{\mbox{\tiny ISM}}$ and particle density
$n_{\mbox{\tiny ISM}}$ of the hot and tenuous outer medium, the energy
density $e_{\mbox{\tiny ISM}}$ of the plasma is given.
The density and temperature profile of the cloud
is then calculated by integrating equations (\ref{gl10}) and (\ref{gl11})
from inside-out
using the core temperature of the cloud as a boundary condition and
truncating the cloud's outer border where the energy density
reaches $e_{\mbox{\tiny ISM}}$.

\section{Models}

Here we present the evolution of three different models.
Their parameters are given in Table~\ref{t2}.
For all models the temperature of the HIM is fixed to $5.6 \cdot 10^{6}$ K.
In two models their density is set to $6.6 \cdot 10^{-4}$ cm$^{-3}$
to allow for comparison with static models of Paper I 
and for one model (E) is increased by one order of magnitude.


      \begin{table*}[t]
      \centering
      \caption{\label{t2}Model parameters used in simulations as described in the
      text. Valid for all simulations: 
      $T_{\mbox{\tiny ISM}}~=~5.6~\cdot~10^{6}$ K i.e. sound velocity $c_{\mbox{\tiny ISM}}       
      = 357$ km s$^{-1}$, $v_{\mbox{\tiny ISM}} = v_{\mbox{\tiny rel}} = 0.3 c_{\mbox{\tiny ISM}}  
      = 107$ km s$^{-1}$} 
      \begin{tabular}{llllllllll}
      \hline
      & $n_{\mbox{\tiny ISM}} / $cm$^{-3}$ 
      & $R_{\mbox{\tiny cl}} / $ pc
      & density
      & $M_{\mbox{\tiny lower}} / $ M$_{\sun}$ 
      & $M_{\mbox{\tiny cl}}    / $ M$_{\sun}$
      & $M_{\mbox{\tiny max}}   / $ M$_{\sun}$ 
      & $\sigma_0$
      & $\lambda_{\mbox{\tiny F}}/R_{\mbox{\tiny cl}}$\\
      Model
      & 
      & 
      & contrast
      & (Eq.~\ref{mlow})
      & 
      & (Eq.~\ref{mmax})
      & (Eq.~\ref{sigma})
      & (Eq.~\ref{field}) \\ \hline \hline
      U & $6.6 \cdot 10^{-4}$ & $41.3$ &
      $1.2 \cdot 10^{4}$ &
      $2.03 \cdot 10^{4}$ & $6.4 \cdot 10^{4}$ & $1.12 \cdot 10^{5}$ &
      $10.5$ & $1.4 \cdot 10^{3}$ \\
      E & $6.6 \cdot 10^{-3}$ & $2.19$ &
      $6.1 \cdot 10^{4}$ &
      $278.7 $ & $486.7$ & $1.3 \cdot 10^{4}$ &
      $19.8$ & $2.6 \cdot 10^{3}$ \\
      K & $6.6 \cdot 10^{-4}$ & $4.0$ &
      $5.9 \cdot 10^{3}$ &
      $3.08 \cdot 10^{5}$ & $15.9$ & $8.3 \cdot 10^{5}$ &
      $108.4$ & $1.4 \cdot 10^{4}$ \\
      \end{tabular}
      \end{table*}


In model U we consider a massive cloud of $6.4 \cdot 10^{4}$
M$_{\sun}$ representing a GMC or PCC. This cloud serves as a
reference model for Paper I in which the fate of this cloud was investigated
in an identical but static environment.
Model E with a cloud mass of $486.7$~M$_{\sun}$ and a cloud
radius one order of magnitude smaller than model U represents
a small molecular cloud. Clouds of this type can be found
as remnants of larger clouds
in galactic chimneys such as the one associated with the
H{\sc ii} region W4 (\cite{he96}; \cite{ta99}) or behind
shock fronts of supernovae.
The size of model K is similar to model E but the mass
is decreased to $15.9$ M$_{\sun}$ to reach an almost homogeneous
density distribution that can be compared in its evaporation rate with
Eq.~(6) which is valid for uniform clouds.
This represents an extreme case of a small cloud that
is only slightly gravitationally bound.

\begin{figure}[ht]
      \psfig{figure=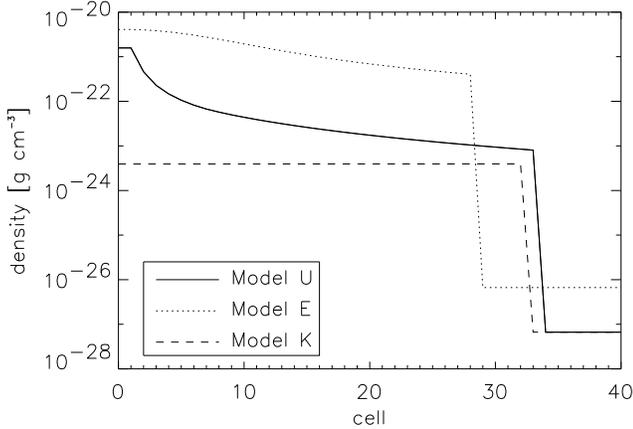,height=6.3cm,width=8.8cm,bbllx=43pt,bblly=396pt,bburx=547pt,bbury=756pt}
      \caption{\label{f1}Density distribution for the various initial models. Model U and E
      show a typical multi-phase structure whereas model K has a uniform density distribution.}
      \end{figure}

All models show a typical multi-phase structure. 
While Model U and E possess a dense core and a density decrease 
outwards, Model K is homogeneous. Their radial density profiles are 
plotted in Fig.~\ref{f1}. 
With respect to dealing with more realistic interstellar clouds with
self-gravity and non-equilibrium boundary conditions, these models 
differ clearly from the situation implied by CM77. 

For comparison all models
were studied with and without heat conduction. The
parameters $\sigma_0$, $\lambda_{\mbox{\tiny F}}$ and
$\tau_{\mbox{\tiny eva}}$ are therefore only defined for calculations
with heat conduction. In all models the hot gas streams
subsonically with $v_{\mbox{\tiny ISM}}$ = 0.3 Mach
and all clouds are kept at rest so that the relative velocity
$v_{\mbox{\tiny rel}}$ important for the KH instability
(see Appendix B) is identical to $v_{\mbox{\tiny ISM}}$.
Such a large velocity difference between clouds and the
HIM is only observed for HVCs, PCCs, and interstellar
clouds overtaken by supernova shocks.
Here we wish to study dynamical effects on clouds in addition 
to heat conduction. For smaller relative velocities, as GMCs 
move through the ISM, the models will approach those of numerical 
heat conduction for clouds at rest in a static HIM (presented 
by us in \cite{vh05}).

The cloud masses were chosen to cover a range
between $10^{1}$ - $10^{5}$~M$_{\sun}$. Because the
initial model has to fulfill Eqs.~(\ref{gl10}) and (\ref{gl11}),
mass and radius are fixed by the initial central
cloud density and external pressure
and temperature.

The saturation parameter $\sigma_0$ (Table~2) indicates a
moderately saturated heat conduction
for model U and E while model K lies nearly in the
suprathermal regime (Balbus \& McKee, \cite{bm82}).
The importance of heat conduction
in these simulations is also
indicated by the fact that the Field length
$\lambda_{\mbox{\tiny F}}$ is much larger than
the cloud radius $R_{\mbox{\tiny cl}}$
which means that evaporation of
the clouds is expected and that the temperature structure of the gas is
dominated by conduction (Begelman \& McKee \cite{bm90}).

The different dynamical timescales $\tau_{\mbox{\tiny dyn}}$ defined as the
sound travel time over one cloud radius, the timescales 
$\tau_{\mbox{\tiny KH}}$
for the growth of KH instability (see Appendix B) and for evaporation are 
listed in Table~\ref{t3}. Because of
the huge density contrast and the large gravitational acceleration of
the models U and E, KH instability should be suppressed, at least at
the beginning. Because of the stability reasons mentioned in 
Appendix B and because
the cloud mass of model E is close to the lower stability limit, this
cloud is expected to be disrupted during the calculation.
Model U is further inside the mass limit than Model E and will therefore
survive longer than the less massive cloud.
Model K on the other hand should
develop KH instability after $\tau_{\mbox{\tiny KH}} = 0.45\,$Myr.

Although the evaporation time of Model U and E is too large to be reached
during the simulation, we should be able to follow the disintegration
process due to evaporation of Model K.

In all calculations of $\tau_{\mbox{\tiny eva}}$ the effect of
saturation is considered.

      \begin{table}[ht]
      \centering
      \caption{\label{t3}Timescales for the three models at their initial
      configuration}
      \begin{tabular}{llll}
      \hline
      & $\tau_{\mbox{\tiny dyn}} = R_{\mbox{\tiny cl}}/c_{\mbox{\tiny cl}}$
      & $\tau_{\mbox{\tiny KH}}$ 
      & $\tau_{\mbox{\tiny eva}}=M/\dot{m}$ \\
      Model
      & Myr
      & Myr
      & Myr \\ \hline \hline
      U & $16.2$ & -- & 279 \\
      E & $1.95$ & -- & 77.1 \\
      K & $0.84$ & 0.45 & 3.56 \\
      \end{tabular}
      \end{table}

      \subsection{Large, massive cloud (model U)}

      \begin{figure*}[ht]
      \psfig{figure=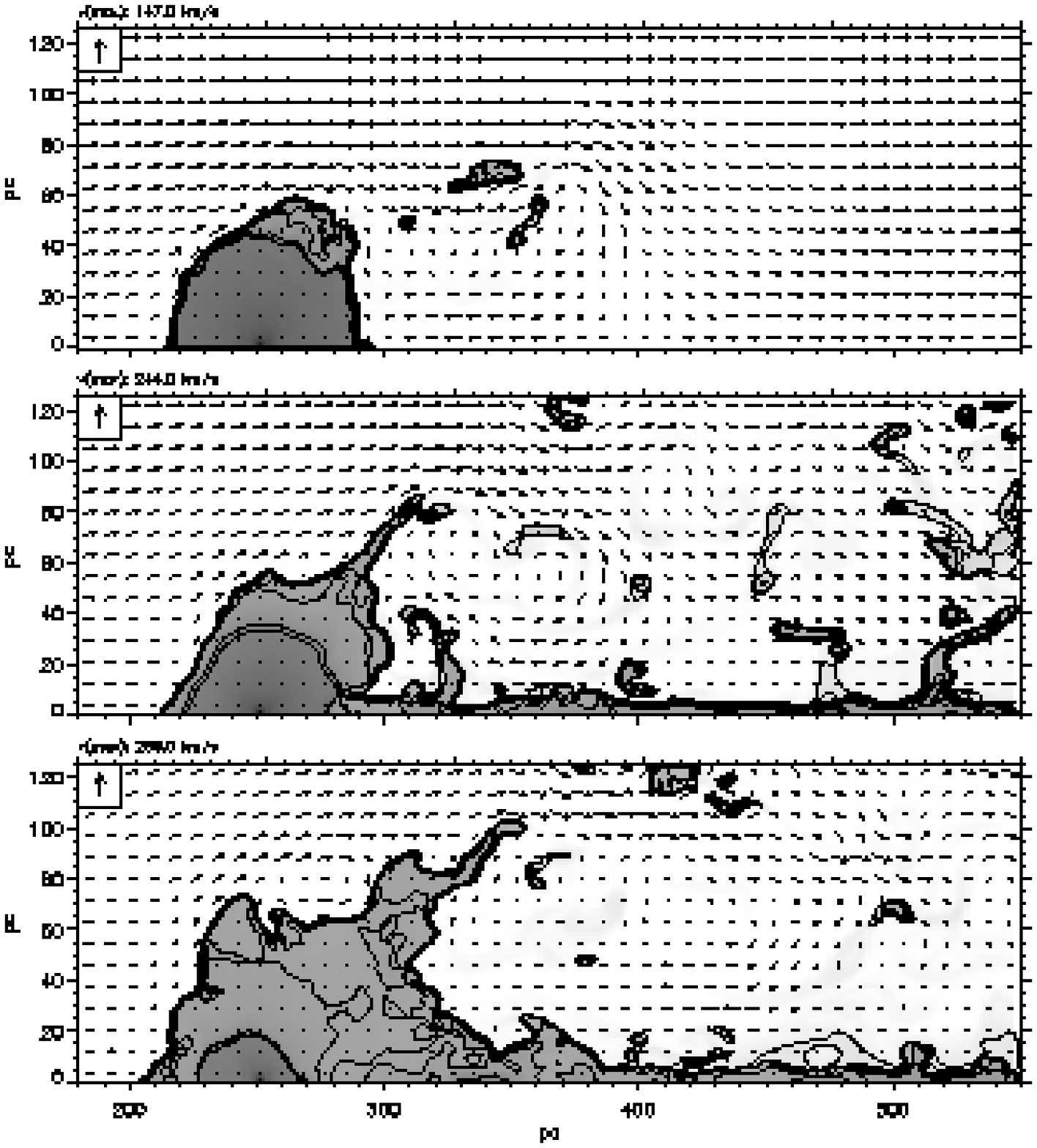,height=18.6cm,width=17cm,bbllx=14pt,bblly=14pt,bburx=511pt,bbury=559pt}
      \caption{
      \label{f2}Evolution of the density distribution for model U
      without heat conduction. The evolution is shown at the times 25 Myr (upper
      panel), 50 Myr (middle) and 75 Myr (lower). Arrows indicate gas
      velocities scaling linearly with respect to the maximum velocity shown in the
      upper left. The greyscale represents the density distribution 
      (logarithmic scale). The contour lines represent 5, 10, 50, 100,\ldots $\times
      \rho_{\mbox{\tiny ISM}}$. The thick contour encloses the gravitationally
      bound part of the cloud.}
      \end{figure*}

      \begin{figure*}[ht]
      \psfig{figure=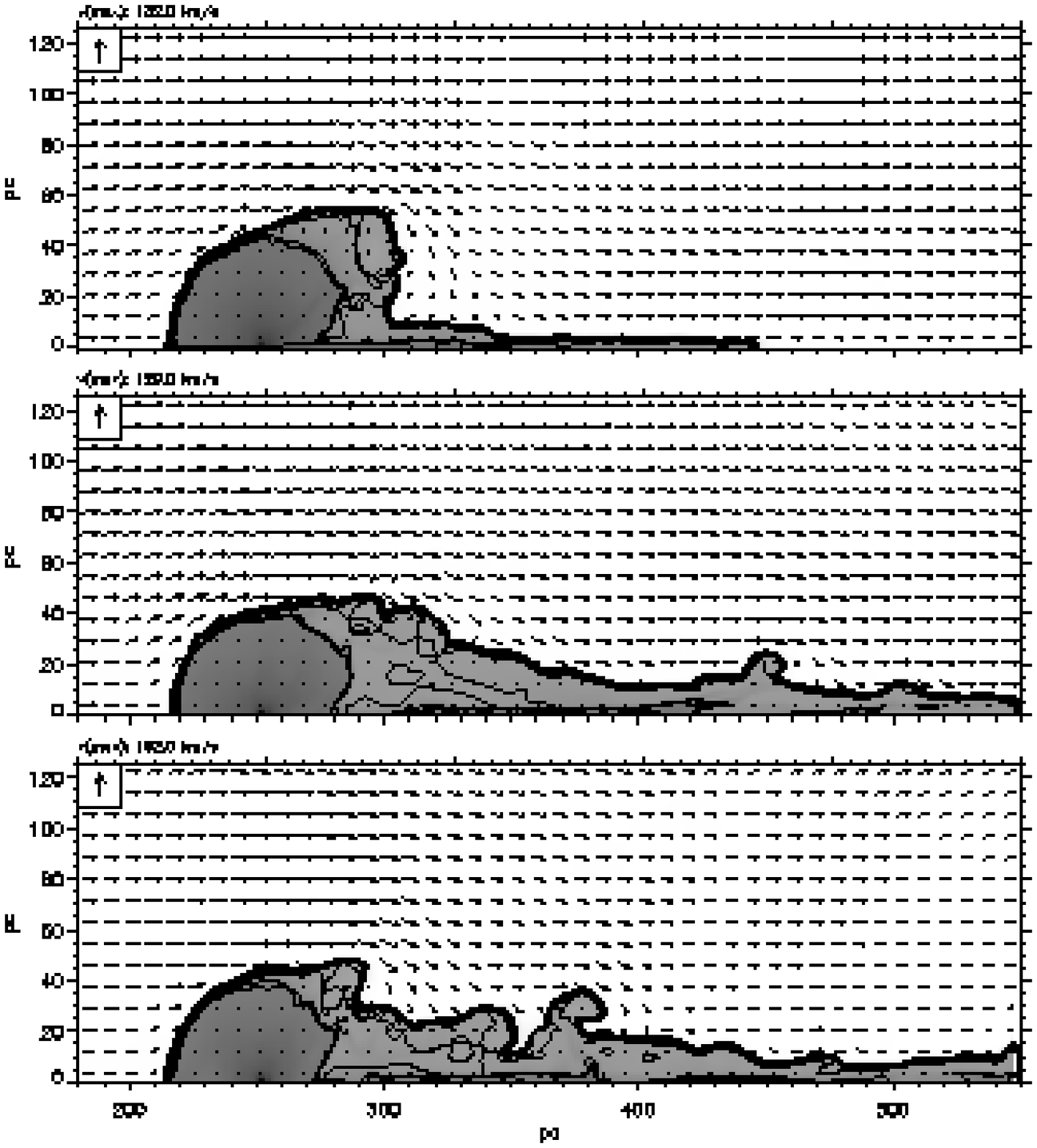,height=18.6cm,width=17cm,bbllx=14pt,bblly=14pt,bburx=511pt,bbury=559pt}
      \caption{\label{f3}Same as Fig.~\ref{f2}: Model U with heat conduction.}
      \end{figure*}

      \begin{figure*}[ht]
      \psfig{figure=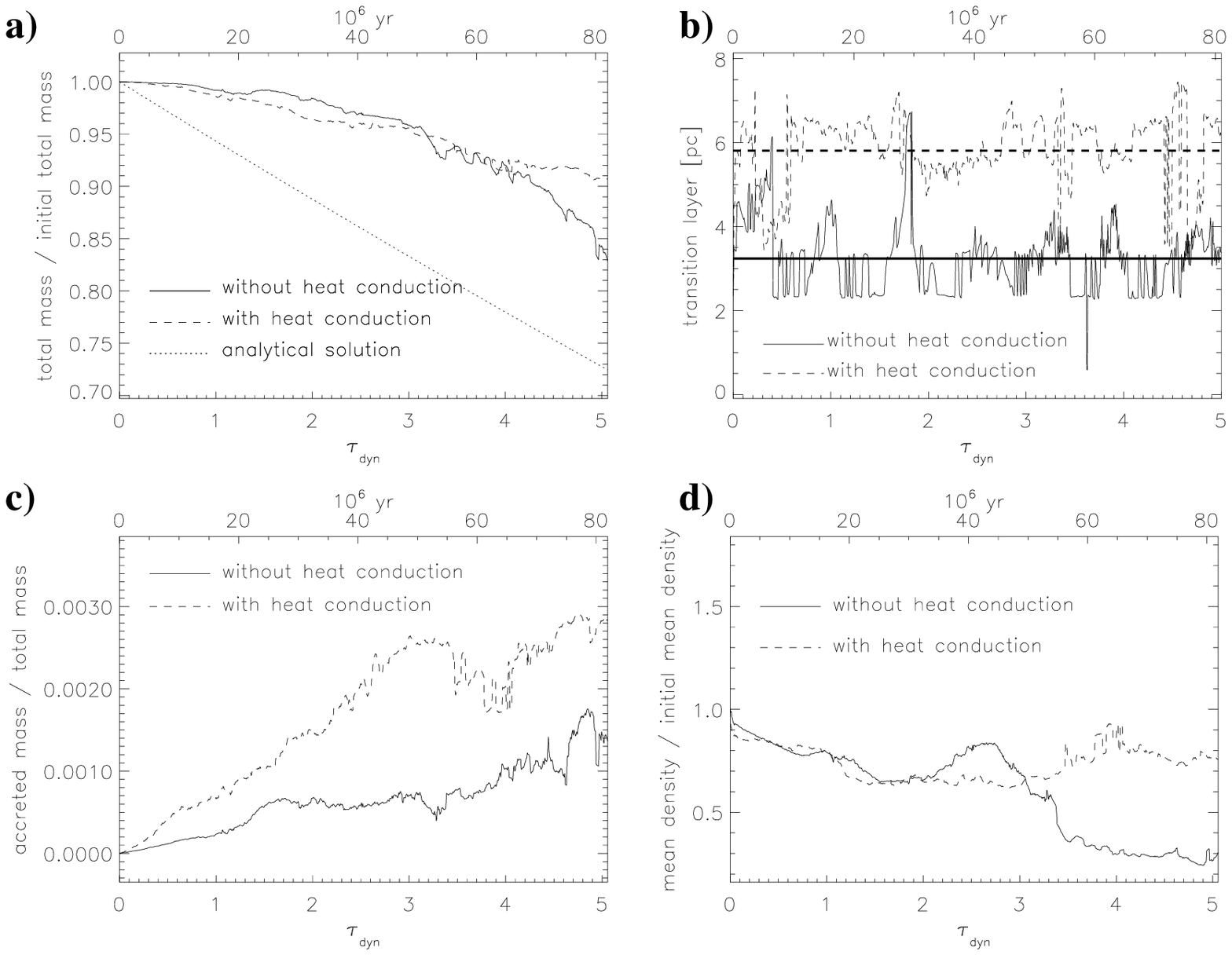,height=13.2cm,width=17cm,bbllx=0pt,bblly=0pt,bburx=481pt,bbury=373pt}
      \caption{\label{f4}Comparison of the evolution of important quantities of model U
      in the case without (solid curve) and with heat conduction (dashed).
      (a) Evolution of the bound cloud mass. The analytical cloud
      mass (dotted line) is calculated taking mass-loss due to evaporation into account.
      (b) Thickness of the transition layer perpendicular to the stream. The thick
      solid and dashed lines represent the mean value for both cases.
      (c) Evolution of the amount of accreted material. (d) Evolution of the mean density.}
      \end{figure*}

The inclusion of heat conduction tends to stabilize massive and
large clouds as illustrated by comparing the density contours at the
same times (25 Myr, 50 Myr and 75 Myr after the beginning of the calculations)
for the model U without (Fig.~\ref{f2}) and with heat conduction
(Fig.~\ref{f3}). Without heat conduction the edge of the cloud is torn
by KH instability shortly after the beginning as mentioned in 
Appendix B.
A very complex velocity structure with many vortices behind the cloud
forms. The increase in the maximum velocity during the calculation
is due to acceleration of the gas by Bernoulli's
effect when it streams around the formed cloudlets. It is therefore only
a local effect which does not affect the large cloud as a whole.
In the presence of cooling stripped-off cloud material becomes thermally
unstable and forms cloudlets that hit the cloud from the rear.
As mentioned by Dinge (\cite{d97}) these cloudlets produce soundwaves which
propagate through the cloud. On the front side these waves produce distortions
that can act as seeds for RT instability that lead to an elongation
of the cloud perpendicular to the stream direction. This effect is
artifically intensified by the applied numerical symmetry. 

Material once collected on the z-axis sticks there and cannot be removed
because the velocity in the radial direction is set to zero at the axis.
A detailed analysis of this artefact and the influence of this axis
effect is given in Appendix A.

The cloud develops a
compound envelope, with a radially decreasing density in the core
region and a diluted outer part of material that is only slightly
gravitationally bound. The loss of gas by stripping and the recapture
from the rear leads to a complex mass-loss function with unsteady
losses and gains. The mass of individual cloudlets that escape into the
ISM lies in the range of some ten solar masses.

Fig.~\ref{f4}a shows the evolution of the gravitationally bound cloud mass 
of the simulations without and with heat conduction. For comparison 
the analytical result of the cloud mass is shown taking mass-loss 
by evaporation into account (Eq.~6).
From this figure we see:
1) The pure evaporation effect on a static homogeneous cloud from CM77
is the most destructive because of its shortest timescale, but it overestimates 
the strength of heat conduction and its heating effect. 
2) The pure dynamical destruction by KH instability happens on
a much longer timescale. As in a static self-gravitating and cooling
cloud model (Paper I), saturated heat conduction yields stabilization 
against evaporation; this more realistic approach also reduces the
dynamical mass loss. While in the static models this is solely caused 
by an increase of heat capacity due to energy transfer to denser regions, 
the dynamical destruction is reduced by weakening the KH effect.
This also means that a cloud that is stabilized in the static case by
heat conduction is ablated solely and less effectively by KH instability.

Obviously the dynamical interaction between the gas phases is the 
dominating effect for cloud destruction in all models. 

After 81 Myr the cloud mass with heat conduction reaches
a value approximately 7\% larger than without,
indicating that heat conduction reduces the ablation of material from the
cloud. A comparison of Fig.~\ref{f2} with Fig.~\ref{f3} shows
significant differences in both the
cloud structure and its evolution. Heat conduction suppresses large-scale
KH instabilities
as indicated by the almost laminar flow pattern around the cloud.
During the whole simulation only a single large circulation
in the slipstream of the cloud is visible. This smoothing process can be
understood by conducting electrons moving along the steepest temperature
gradient, i.e.
radially toward the cloud, so that the stream close to the contact
interface is decelerated, the velocity gradient decreases and the
transition layer (described in Appendix B) grows. Numerically, 
the transition layer is by definition the region in which the 
velocities transit from the HIM to the clouds' thermal value. 

Fig.~\ref{f4}b shows that over the entire simulation time the thickness 
of the transition layer perpendicular to the stream is, in the heat conduction 
case, nearly twice as thick as in the non-conductive case. 
Comparison with Fig.~\ref{sb2} reveals that the transition layer
without heat conduction lies in the instability zone while it approaches
or even exceeds the critical layer thickness d for stability. This effect 
is also responsible for the more efficient accretion and incorporation of 
streaming material into the cloud visible in Fig.~\ref{f4}c. The lower 
velocities in the contact interface enable the streaming hot ISM to condense
onto the cloud surface. Thereafter it is very efficiently mixed with
cloud material and transported into deeper layers of the cloud by turbulent 
motion. This turbulence is also driven by heat conduction that acts as an
additional energy source in low-density regions where it exceeds the cooling.
Without heat conduction the accreted material is peeled off from the surface
by the grazing high-velocity gas before it could be incorporated into
the cloud.

The multi-phase cloud in our simulation can be divided into two
parts (see Fig.~5): 1) regions smaller than $\lambda_{\mbox{\tiny F}}$
where the heat conduction exceeds the cooling so that the density
distribution becomes homogenized; 2) very dense regions, such as near
the core where the density structure is unaffected by heat conduction.
A further surprising result from the numerical models is 
that the heat conduction becomes less effective after the
low-density outermost shells of the cloud are dissolved.
The resistance of the major part of the cloud against the disruptive
KH instability and evaporation processes can be revealed from the
temporal behaviour of
the mean density of the bound mass (see Fig.~\ref{f4}d). Until $2.5
\tau_{\mbox{\tiny dyn}}$ the decline of the mean density is due
to the dissolution of the outer parts of the cloud.
Thereafter the mean density increases again and remains
nearly constant at a value of around 80\% of the initial mean density.
In contrast the decrease of the mean density in the model without
heat conduction shows the continuous dispersion
of the whole cloud by lowering the density also in the core regions.
This leads to a flat gravitational potential which further destabilizes 
the cloud.

\subsection{Dense cloud (model E)}

\begin{figure*}[ht]
      \psfig{figure=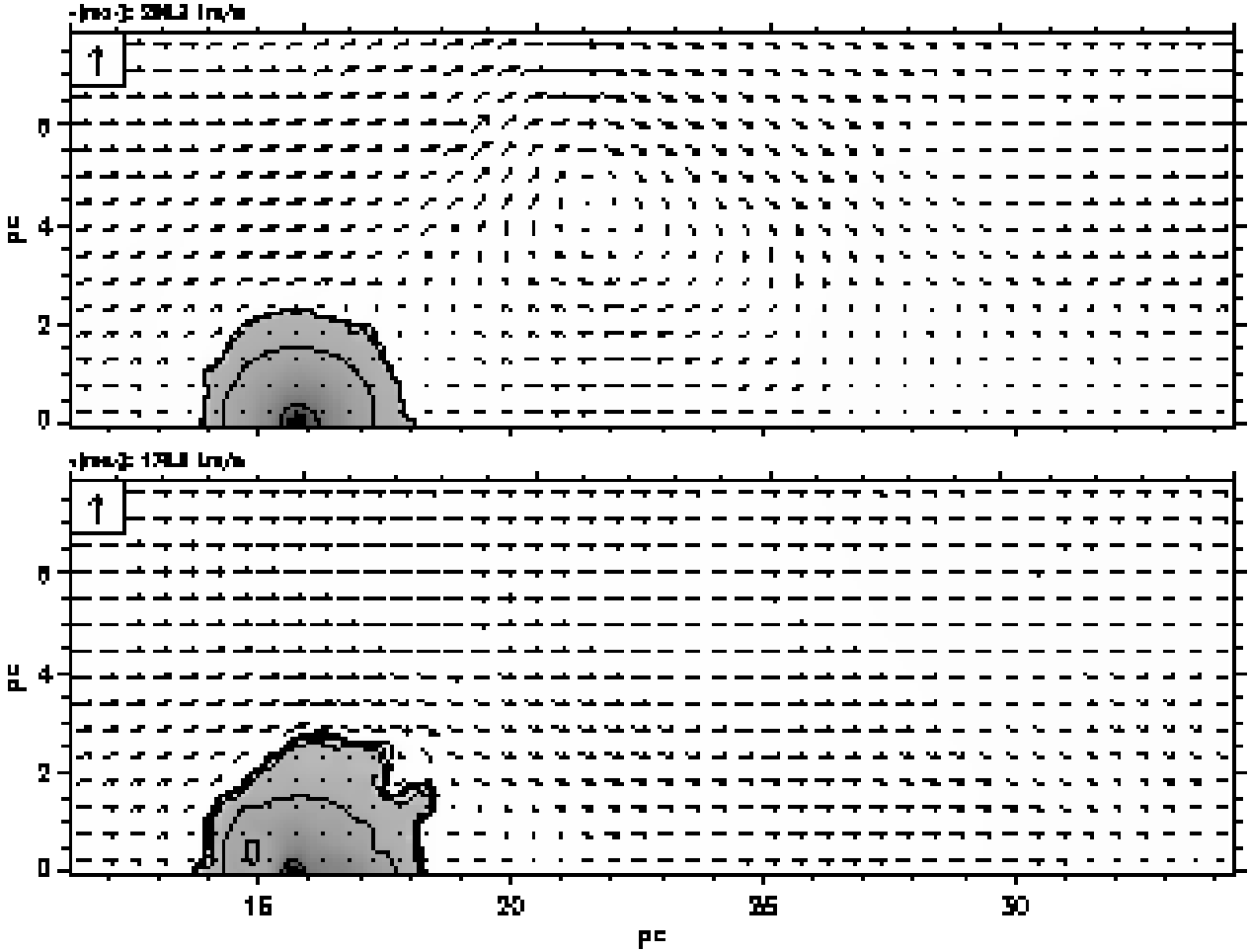,height=13cm,width=17cm,bbllx=14pt,bblly=14pt,bburx=491pt,bbury=382pt}
      \caption{\label{f7}Snapshot of the density distribution for the simulation E
      without heat conduction (upper part) and with heat conduction (lower part). 
      The evolution is shown at 22.5 Myr. Arrows indicate gas
      velocities scaling linearly with respect to the maximum velocity shown in the
      upper left. The greyscale represents the density distribution 
      (logarithmic scale). The contour lines represent 5, 10, 50, 100,\ldots $\times
      \rho_{\mbox{\tiny ISM}}$. The thick contour encloses the gravitationally
      bound part of the cloud.}
      \end{figure*}

To investigate the effect of heat conduction on denser, 
less massive, i.e. more compact clouds we increase the external density 
$\rho_{\mbox{\tiny ISM}}$ by one order of magnitude. 
Because the evolution of this model is not as violent as the previous
one, a snapshot of the evolution 22.5 Myr after the begining of the 
calculations is shown in Fig.~\ref{f7} without (upper part) and with
heat conduction (lower part) for comparison.

      \begin{figure*}[ht]
      \psfig{figure=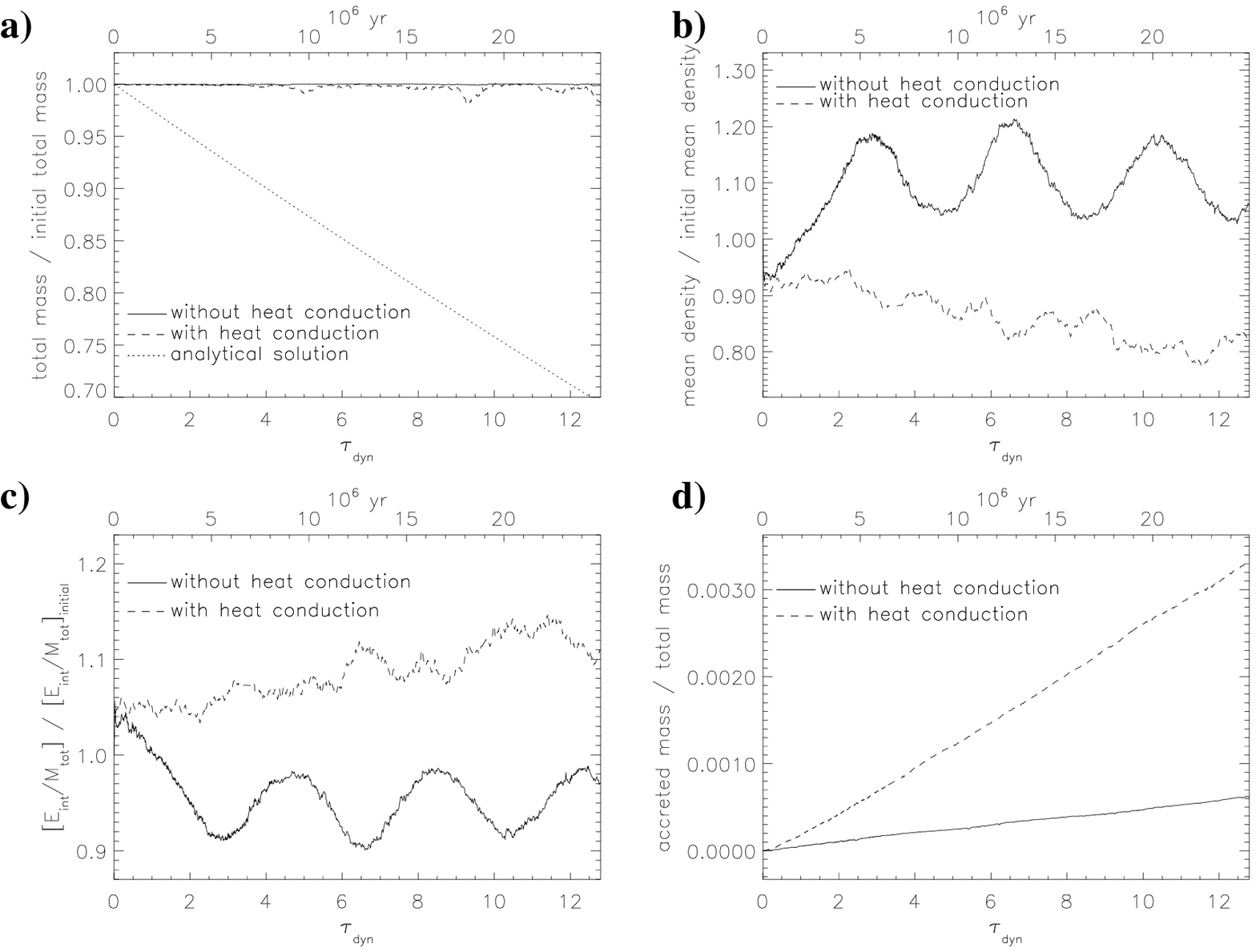,height=12.8cm,width=17cm,bbllx=0pt,bblly=0pt,bburx=494pt,bbury=373pt}
      \caption{\label{f9}Comparison of the evolution of important quantities of
      model E in the case without (solid curve) and with heat conduction (dashed).
      (a) Evolution of the bound cloud mass. The analytical cloud
      mass (dotted line) is calculated taking mass-loss due to evaporation into account.
      (b) Evolution of the mean density. (c) Evolution of the specific internal
      energy. (d) Evolution of the amount of accreted material.}
      \end{figure*}

In the non-conductive case even after 22.5 Myr no KH instability is visible 
because of the stabilizing effect by the high gravitational 
acceleration at the cloud surface.
 
While the shape of the cloud remains unaltered, variations of the
mean cloud density and the specific internal energy are present 
in Fig.~\ref{f9}c. These pulsations stem from quasiadiabatic oscillations 
with a timescale given by
$\tau_{\mbox{\tiny ad}}=\sqrt{ 3 \pi/[ (3 \gamma - 4)\rm{G} \bar{\rho}] }$. 
$\bar{\rho}$ is the mean cloud density.

For model E and the adiabatic index $\gamma = 5/3$ ,
$\tau_{\mbox{\tiny ad}}$ is 7 Myr which agrees well
with the obtained one.
These oscillations originate from the interaction of the plasma stream
with the cloud at the beginning of the simulation. This process triggers
accoustic waves that run through the cloud that lead locally to
density enhancements. 

In addition, the static models (\cite{vh05}) demonstrate that
heat conduction produces inward travelling sound waves which are able to 
transport energy from the conductive interface to the denser cloud interior. 
By this additional effect the cloud releases the transferred energy.

In the case with heat conduction, small-scale mixing at the cloud
surface with the outer HIM leads to an increase of the cloud volume
visible as a decrease in the mean density (Fig.~\ref{f9}b).
The oscillations occurring in the non-conductive case are efficiently
damped by turbulent motions of the cloud material driven by heat conduction.
The incorporation of hot material into the cloud acts as an additional
heating agent that raises the temperature and therefore
the pressure of the cloud and results in an expansion. This
temperature increase is reflected by the increase of the specific 
internal energy (Fig.~\ref{f9}c). 
Regions of the cloud near the vortex in the slipstream
are especially heated and diluted because of the more efficient
mixing there. 
This material is only slightly gravitationally bound and 
therefore preferentially stripped off the cloud which can be revealed 
by the distorted cloud shape in those regions. Also the small mass-loss
of 2\% at about 25 Myr (12.8 $\tau_{\mbox{\tiny dyn}}$) is caused
by this effect. The distortion of the cloud to a more streamlined
shape as visible in the lower part of Fig.~\ref{f7} 
leads the flow around the cloud in 
a more laminar manner.

As in model U the amount of
accreted material is again higher with heat conduction than without
(Fig.~\ref{f9}d). This is caused by the suppression of the
velocity differences in the transition layer. However,
in the conductive model the absolute accretion rate normalized
to the total mass is much higher here on absolute timescales
(after 20 Myr e.g. 0.3\% are accreted in model E at constant rate
but almost 0.1\% in model U) but lower on $\tau_{\mbox{\tiny dyn}}$
( 0.26\% within 10 $\tau_{\mbox{\tiny dyn,E}}$ in model E, 
while 0.25 \% within 3 $\tau_{\mbox{\tiny dyn,U}}$ in model U).

In order to understand this effect quantitatively for the different 
parameters of the two models one can make a zeroth-order 
approach neglecting density and mass changes. 
An accretion rate $\Omega$ can be 
defined as $M_{\mbox{\tiny acc}}$ per dynamical time 
$\tau_{\mbox{\tiny dyn}}$ and reformulated by normalization to
the total mass $M_{\mbox{\tiny tot}}$ as shown for model U in 
Fig.\ref{f4}c and for model E in Fig.\ref{f9}d, respectively, by

      \begin{equation}{\label{o0}}
      \Omega = {\frac{M_{\mbox{\tiny acc}}}{M_{\mbox{\tiny tot}}}} \; 
               M_{\mbox{\tiny tot}} \; \tau_{\mbox{\tiny dyn}}^{-1}
      \end{equation}

where $\tau_{\mbox{\tiny dyn}}$ must not be further determined. 
Applying this relation, the ratio 
$\Omega_{\mbox{\tiny U}}/\Omega_{\mbox{\tiny E}}$ follows
as

      \begin{equation}{\label{o1}}
      {\frac{\Omega_{\mbox{\tiny U}}}{\Omega_{\mbox{\tiny E}}}} = 
      ( {\frac{0.0025 \; M_{\mbox{\tiny cl,U}}}{3 \; \tau_{\mbox{\tiny dyn,U}}} })/
      ( {\frac{0.0026 \; M_{\mbox{\tiny cl,E}}}{10 \; \tau_{\mbox{\tiny dyn,E}}} }) \; .
      \end{equation}

With Eq.~(\ref{o0}), 
$M_{\mbox{\tiny acc}} = \Omega \; \tau_{\mbox{\tiny dyn}}$
follows so that one can write the ratio 
$M_{\mbox{\tiny acc,U}}/M_{\mbox{\tiny acc,E}}$
where the different $\tau_{\mbox{\tiny dyn}}$ cancel out with 
Eq.~(\ref{o1}) and one obtains the value 421.  
This value agrees as an order-of-magnitude effect within the 
uncertainties and being aware of the simple estimates of the 
structure, if one assumes that the condensation 
$M_{\mbox{\tiny acc}}$ depends on the cloud's surface, 
the surrounding density and the Field length is given by:

      \begin{equation}
      M_{\mbox{\tiny acc}} \; \propto \; 
          \lambda_{\mbox{\tiny F}} \; R_{\mbox{\tiny cl}}^2 \; 
              n_{\mbox{\tiny ISM}} = 
          \frac{\lambda_{\mbox{\tiny F}}}{R_{\mbox{\tiny cl}}} \;
              R_{\mbox{\tiny cl}}^3 \; n_{\mbox{\tiny ISM}}  \;  .
      \end{equation}

This ratio for models U and E is 361.

Compared to model U where heat conduction prevents
the disruption of the cloud, in this model the cloud is already stabilized
without heat conduction and condensation dominates the evolution.
The dip in the mass loss curve (Fig.~9a) after 18 Myr is caused by the
separation
of a small cloudlet with successive rebounce so that it remains invisible
in Fig.~9d. Evaporation, however, is then not directly caused by heat conduction.
Instead, conduction expands the cloud and KH instabilities dissolve
the back parts of the cloud.
Fig.~\ref{f9}a reveals that the total bound mass remains constant for the case
without heat conduction but decreases with heat conduction only by 2\%
within 12.8 $\tau_{\mbox{\tiny dyn}}$.
Interestingly both conductive models U and E experience an expansion that
reduces the mean density by nearly 20 \%.
Although heat conduction leads to
a slight expansion of the cloud and, by this, to this minor fraction of mass
loss, the analytical result in contrast requires evaporation of
30\% of the cloud after 12.8 $\tau_{\mbox{\tiny dyn}}$. 
The analytical prescription overestimates the
real effect by far, when both gas phases are dynamically different.

\subsection{Small, homogeneous cloud (model K)}

\begin{figure*}[ht]
      \psfig{figure=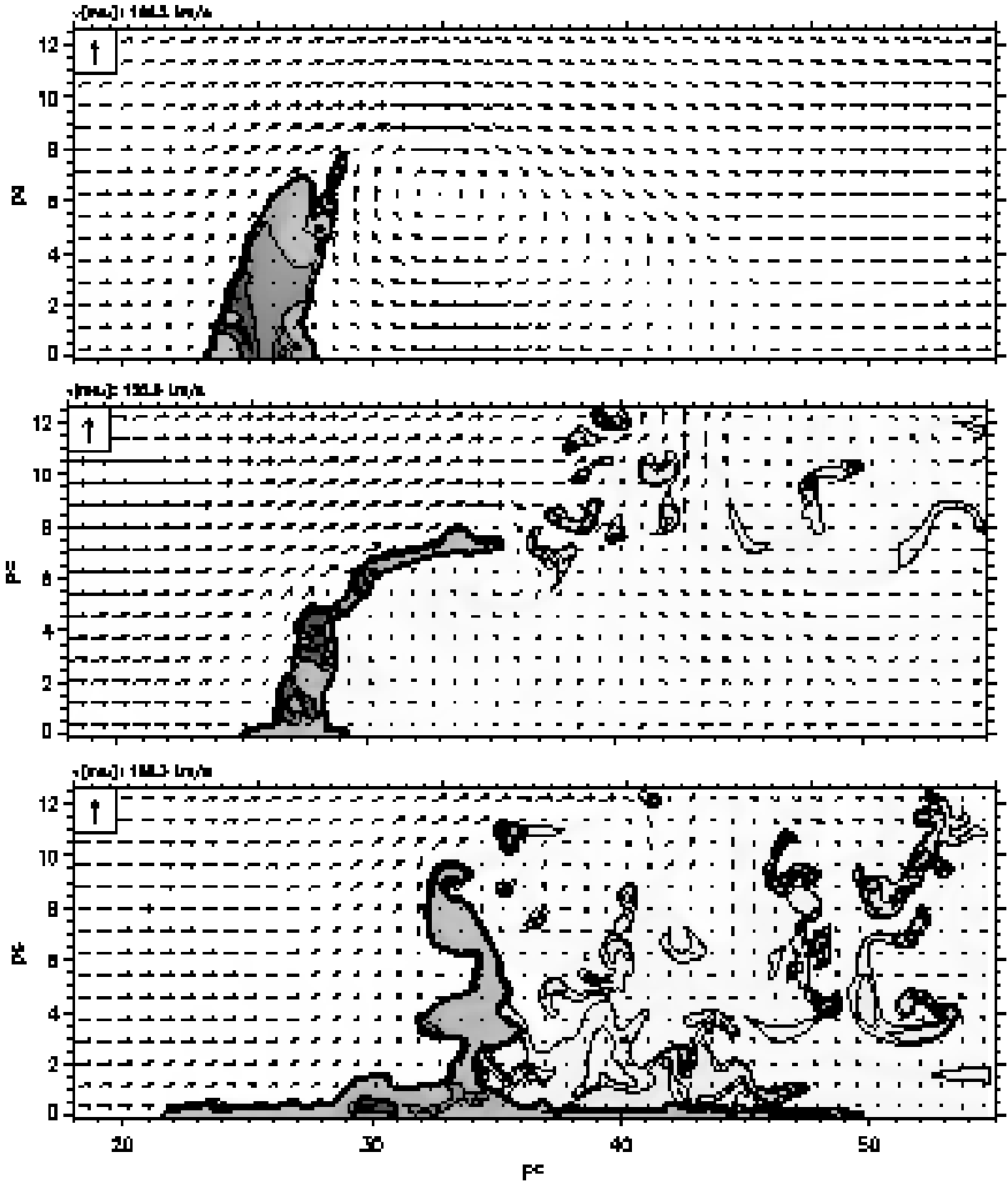,height=19.2cm,width=16.2cm,bbllx=14pt,bblly=14pt,bburx=500pt,bbury=562pt}
      \caption{\label{f11} Evolution of the density distribution for the simulation K
      without heat conduction. The evolution is shown at the times 3.4 Myr (upper
      panel), 6.8 Myr (middle) and 10.2 Myr (lower). Arrows indicate gas
      velocities scaling linearly with respect to the maximum velocity shown in the
      upper left. The greyscale represents the density distribution (logarithmic scale). 
      The contour lines represent 5, 10, 50, 100,\ldots $\times
      \rho_{\mbox{\tiny ISM}}$. The thick contour encloses the
      the domain where $\rho_{\mbox{\tiny cl}} > 100
      \times \rho_{\mbox{\tiny ISM}}$.}
      \end{figure*}

\begin{figure*}[ht]
      \psfig{figure=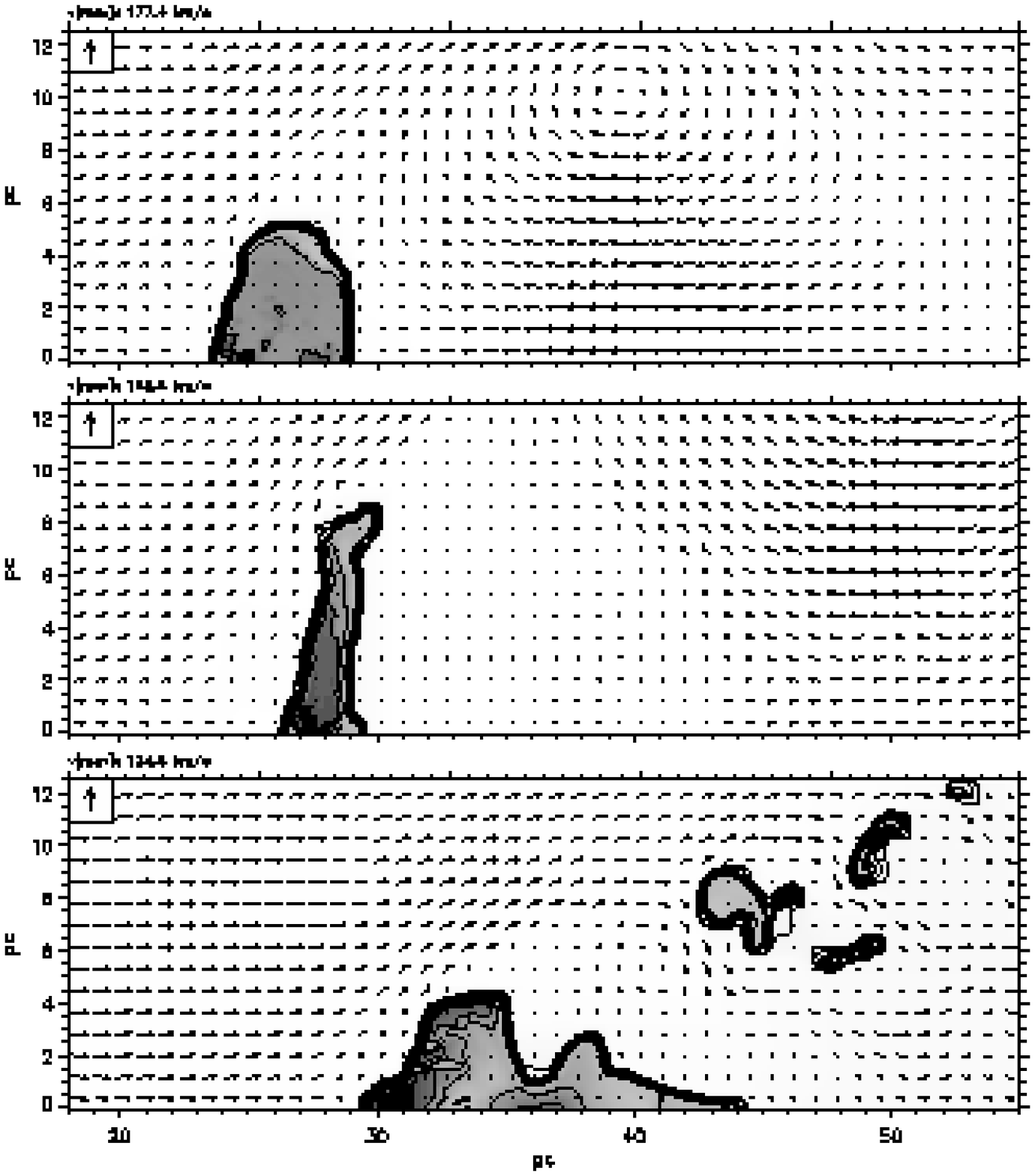,height=19.2cm,width=16.2cm,bbllx=14pt,bblly=14pt,bburx=497pt,bbury=562pt}
      \caption{
      \label{f12}Same as Fig.~\ref{f9}: Model K with heat conduction.}
      \end{figure*}

\begin{figure*}[ht]
      \psfig{figure=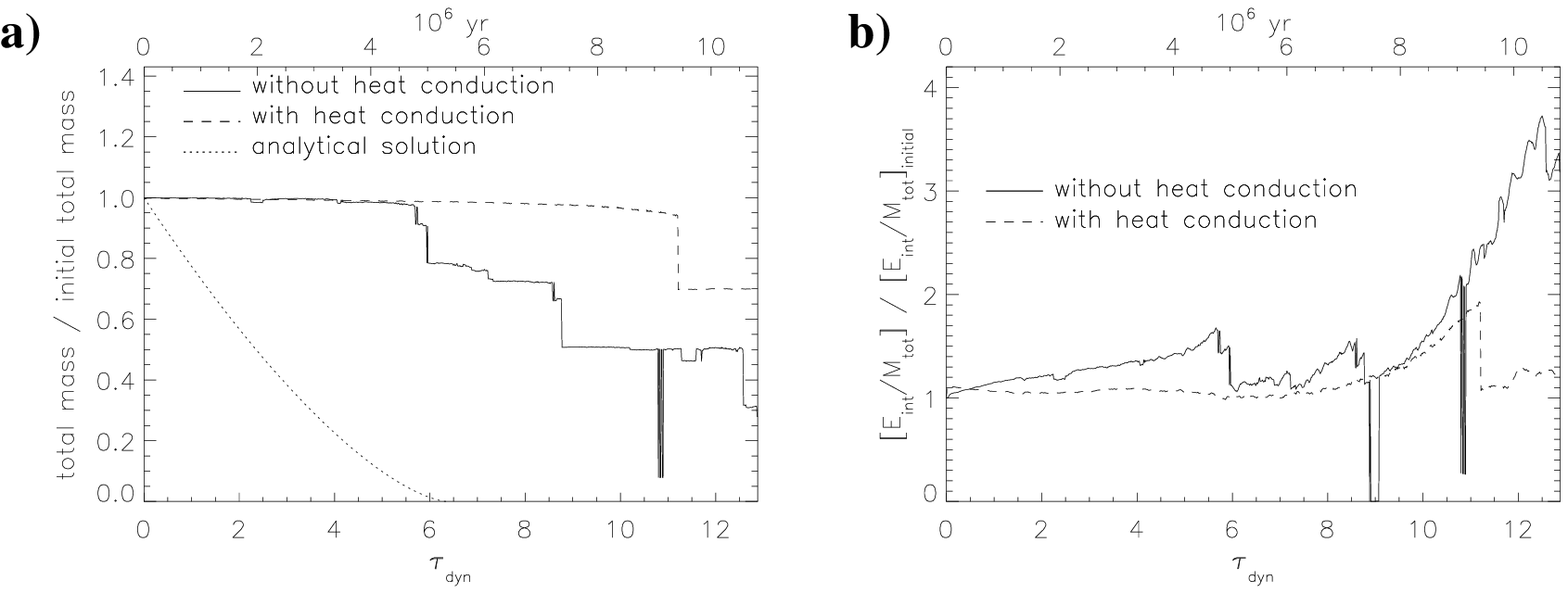,height=6.4cm,width=17cm,bbllx=0pt,bblly=0pt,bburx=489pt,bbury=184pt}
      \caption{\label{f13}
      Comparison of the evolution of important quantities of model K in the
      case without (solid curve) and with heat conduction (dashed).
      (a) Evolution of the bound cloud mass. The analytical cloud mass (dotted line)
      is calculated taking mass loss due to evaporation into account.
      (b) Evolution of the specific internal energy.}
      \end{figure*}

A third model describes an extreme case:
Model K represents a small cloud with no distinct core 
and a density contrast of almost one order of magnitude lower
than the stable model E. The analytical consideration of the 
classical heat conduction (CM77), therefore, requires fast and 
efficient evaporation (see Table~\ref{t3} and Fig.~\ref{f13}b). This cloud 
is only marginally gravitationally bound, so that its gravitational 
acceleration cannot sufficiently stabilize it against KH instability 
(see Appendix B).
The timescale for the onset of KH instability is
approximately $\tau_{\mbox{\tiny KH}} =$ 0.45 Myr, i.e. almost half
of $\tau_{\mbox{\tiny dyn}}$. 
The cloud should therefore be destroyed within a few dynamical timescales.
The violent action of KH instability is easily visible in Fig.~\ref{f11}
where the
density contours for Model K without heat conduction
are shown in steps of
3.4 Myr after the beginning of the calculations.
The density contours in the case with heat conduction
are revealed in Fig.~\ref{f12} for the same timesteps.

Fig.~\ref{f11} shows that the cloud model K is already very fragile at
the beginning of the simulation. The cloud is soon perpendicularly
stretched because of the low pressure at both sides due to Bernoulli's 
theorem. The density in these regions is lowered and its cooling ability 
diminished. This, together with the mixing of hot ISM into the cloud
due to small scale perturbations triggered by means of 
KH instability leads to an increase of the temperature
which can be discerned as an increase of the specific internal energy
(Fig.~\ref{f13}b). This energy input is sufficient to unbind the
cloud material.
From Fig.~\ref{f13}b one also sees that the stripped-off 
gas contains a larger specific internal energy than the rest, i.e.\ 
energy that is acquired from dynamically excited turbulence is lost.
Due to the initiated turbulent velocity field inside the cloud the 
mass distribution changes throughout the calculation.
Consequently, no distinct core can be 
formed. The low binding energy allows the cloud material to be
caught by the gas flow and to be carried away at about
5 Myr (5.9 $\tau_{\mbox{\tiny dyn}}$). Because the stripped cloud 
material is disrupted into small packages, the mass-loss rate  
(Fig.~\ref{f13}a) shows a step-wise function.   
After about 10.8 Myr (12.8 $\tau_{\mbox{\tiny dyn}}$) the cloud is nearly
destroyed and only 30\% of the initial cloud mass has survived.

The evolution in the case with heat conduction is very similar during 
the first 5 Myr to that without heat conduction. Both the elongation of the cloud
and the mass-loss rate show nearly the same behaviour. In contrast to the
non-conductive case, KH instability is suppressed and thus the additional
heat input through mixing with the hot ISM is reduced. The temperature and
therefore the specific internal energy rises only moderately (Fig.~\ref{f13}b).
Consequently, the velocity field within the cloud is less turbulent
so that a core can form. The
cloud interior develops a relatively stable mass distribution with a
dense core which stabilizes this model against mass loss at least twice as long
(Fig.~\ref{f13}a). Due to heat conduction a transition layer with a
radial decrease of the density and velocity gradient has formed and with this
a radial temperature gradient along which the cloud material flows.
Some material is lost after 7.5 Myr when a large cloudlet with a mass
of about 30\% of the whole cloud decouples from the elongated cloud
at its top.

As in large massive clouds (model U) where the development of a
transition zone near the cloud surface leads to a reduction of KH
instability, the same is valid for small homogeneous clouds.
No large-scale KH instability is seen although the initial state of 
the cloud is very unstable against it.
At the end of the calculation the total mass of
the cloud is about 30\% higher in the case with heat conduction than
without. 

The heat conduction process leads to 
  dynamical processes at the clouds' surface that decelerates the disintegration
  of the whole cloud by suppressing the large scale KH instability. In
  addition, considering cooling in the heat conduction description stabilizes
  this cloud against mass loss, while the analytical
  solution would again destroy the cloud by evaporation much more rapidly,
  within 6 $\tau_{\mbox{\tiny dyn}}$.

\section{Discussion}

In agreement with analytical results for the static case (CM77)
we have proven that heat conduction must not be neglected
in investigations of the evolution of interstellar clouds in a hot dilute
plasma and, in general, of the coexistence of the multi-phase ISM.
While \cite{cm77} indicate that all the clouds in this paper would 
undergo evaporation due to heat conduction, we 
demonstrated in \cite{vh05} and in sect. 1.3
that a realistic approach that accounts for the electron conductivity
properly leads to the opposite results, namely, that condensation dominates.
The cloud conditions applied to our models differ from the consideration of CM77.
In this paper we have shown with models
that the fate of a cloud and its evolution is drastically changed 
by a gas stream and depend strongly on the mass and its internal structure 
of the cloud, i.e. mainly on its binding energy.
In general, the relative motion of a subsonically hot plasma stream
stabilizes the clouds.

Without heat conduction, clouds with their initial states close to or
inside the KH-unstable regime suffer from huge mass loss in the form
of stripped-off cloudlets. While a small homogeneous cloud (model K)
is stable for about 5 $\tau_{\mbox{\tiny dyn}}$ and then strongly exposed
to disruption into small gas packets,
small dense clouds (model E) can avoid the transition into the KH-unstable
state and resist the violent hot plasma
so that mass loss or even any strong deformation of the cloud does not occur.
Large massive clouds (model U) lose about 17\% of their mass within
5 $\tau_{\mbox{\tiny dyn}}$ and may dissolve on larger timescales.

For the static case of the coexistence of two gas phases, namely, clouds
embedded in a HIM, the mass transfer according to heat conduction
can be calculated analytically (CM77). We have chosen
plasma conditions that would require evaporation of cloud material and 
therefore mass loss from the cloud from analytical considerationsly. 
Under the dynamical action of a relative motion between the gas phases the
conditions change with respect to the static case in two ways:
Dynamical instabilities can change the shape of the cloud and can increase
its surface. In contrast to the analytical results, the state of the
hot ISM remains constant because of its replenishment by the fixed
streaming conditions and therefore cannot react to the evaporation and
condensation process and by this e.g. self-regulate the mass
transfer to find an equilibrium (see e.g. K\"oppen et al. 1998).

Large and massive clouds survive longer in the hot plasma flow with heat 
conduction than without. Because of electron invasion through the surface, 
a transition zone forms at the edge of the cloud where density and velocity 
gradients are lowered.
A state can be reached where the KH instability is suppressed and the 
formerly unstable cloud becomes stabilized. This can be shown analytically 
and numerically reproduced (model U).
Since the evaporation rate is much less than the one
predicted by \cite{cm77} these facts lead to a cloud mass at the
end of the calculation that is even slightly higher (7\%) than in the case
without heat conduction. 

Although the maximum cloud mass implied here is much lower than those
of HVC complexes moving through the galactic halo with masses of
a few $10^6$ M$_{\sun}$ such as Complex C (Wakker et al. \cite{w88})
or compact HVCs located in the intergalactic medium (\cite{bb99}),
only clumpy substructures seem to decouple from complexes and approach 
the galactic disk and experience on their path through the halo interaction 
with the hot gas that leads to the observed head-tail structures (\cite{br00}).
Their masses, on the other hand, range from a few umpteen solar masses to a 
few 10$^4$ M$_{\sun}$, like recently found compact HVCs in the inner galaxy 
(\cite{st06}). 
Nevertheless, calculations with even higher masses are in preparation but 
it can be expected that the tendency to stabilize the cloud and to reduce 
the ablation of material from the cloud will continue.
Heat conduction is therefore a physical process that enhances the dynamical
stabilization and has to be taken into account in the consideration
of the survival of HVCs.

For the PCCs, heat conduction offers a mechanism to incorporate metal-rich
hot gas that becomes homogeneously mixed. Even with a low accretion fraction
of only 0.3\% a hot gas metal content of solar and above, which is 
reasonable e.g.\ from X-ray determinations of the halo gas around giant ellipticals
(\cite{ma03}), would lead to almost 1/100 Z$_{\sun}$. Globular clusters
formed from PCCs and enriched by this accretion mechanism caused by heat 
conduction must be expected to show a large range of metallicities. 

When star formation sets in, all protostars are formed from molecular clouds with
nearly equal metallicity. The absence of a significant spread in [Fe/H]
in most globular clusters (Freeman \& Norris \cite{fn81};
Fahlmann et al. \cite{frv85}; Norris \cite{n88}) is an
indication that the stars have formed
out of well mixed metal-enriched substrates that could have been polluted
by an external source (Murray \& Lin \cite{ml90}). This mechanism
provides an alternative explanation to the self-enrichment scenario of
globular clusters (\cite{br91}, \cite{br95}).

When looking at smaller clouds one has to distinguish between a homogeneous 
(model K) and a clearly centrally peaked density distribution (model E).
In the latter case the gravitational potential is strong enough to stabilize 
the cloud against large-scale perturbations triggered by the dynamical
action of the streaming ISM. The influence of the HIM is limited to
an additional heat input due to heat conduction and small-scale mixing
especially in regions near the vortex in the slipstream of the cloud.
Because of the high density in the core regions, the additional heat
input is nearly compensated by cooling. Only the rim of the cloud
becomes diluted and so susceptible to evaporation. Thus, the mass loss 
(2\%) is reduced by more than a factor of 15 with respect to the analytical
approach ($\sim 30$\%). In this case the fate of the cloud depends only on
small-scale, not on large-scale dynamics.

These results cannot be extrapolated to homogeneous clouds (model K), where
already a small additional heat input is sufficient to unbind the whole cloud.
Although KH instability is also suppressed in this case, the cloud becomes
elongated by the Bernoulli effect, which is very efficient because of
the flat gravitational potential. It is this large-scale
phenomenon that peels off a large fragment of the cloud. Evaporation
due to heat conduction is negligible in comparison to the mass loss due to
the large-scale effects. Heat conduction acts instead as an agent to
extend the survival time of the cloud by the suppression of
KH instability and therefore the reduction of the mass-loss
in comparison to the non-conductive case.
While the analytical approach of model K leads to cloud destruction
by evaporation within almost 6 $\tau_{\mbox{\tiny dyn}}$ (Fig.~\ref{f13}a),
in the numerical simulation the dynamics stabilize the cloud and heat
conduction extends the destruction time to more than
11 $\tau_{\mbox{\tiny dyn}}$. After the loss of a large fragment with
30\% of the cloud mass, the total disruption
of the cloud is most likely a consequence of the
dynamics of the streaming ISM. It is therefore not surprising that the
analytical evaporation rates of CM77 are incompatible with our calculations.

Comparing the \cite{cm77} evaporation rates with computed models
for the static case (\cite{vh05}) reveal that condensation may occur for
large clouds in temperature regimes where \cite{cm77} also predict evaporation.

The three presented models can only give a first insight
into the fundamental importance of dynamics and additional heat conduction
in the evolution of the ISM phases. Further investigations are necessary to 
determine the dependence of evaporation and condensation
on the physical state of the phases. This is necessary for the simulation
of galaxy evolution (see e.g. Samland et al. 1997, Hensler 1999).

The effect of magnetic fields on 
heat conduction must be discussed, especially when it is neglected, 
as in our models. When electrons of the HIM enter an interstellar cloud 
they transfer their energy to the cloud by collisions with mainly 
neutral HI atoms of density n$_{\rm HI}$ and collisional cross section 
q$_{\rm HI}$. Their collisional mean free path 
$\lambda_{c,{\rm HI}}$/cm = (n$_{\rm HI} \cdot$ q$_{\rm HI})^{-1}$ 
is about 10$^{16}$ cm. 
On the other hand, magnetic fields force charged particles with mass m$_e$ 
moving with velocity $u_{e,ICM}$ perpendicular to the $B$ field vectors due 
to the Lorentz force to gyrate with the so-called (Larmor) radius a$_{L,e}$. 
For electrons this reads as

\begin{equation}
{\rm a}_{L,e} = \frac{c}{|e|} \frac{m_e \cdot u_{e,ICM}}{B} \; .
\end{equation}

Since we assume in our modelling that at the clouds' surface, both gas phases 
achieve pressure equilibrium we can also set the 
magnetic pressure P$_{mag}$ = $B^2/8\pi$ in equilibrium. 
Replacing $m_e\cdot u_{e,ICM}^2$ by the ICM temperature T$_{ICM}$, it cancels 
with that of the pressure and one gets 
a$_{L,e} = 4.46\cdot 10^5 {\rm n}^{-1/2}_{{\rm H},ICM}$. 

The ratio of mean free path to electron Larmor radius then is

\begin{equation}
\frac{{\rm a}_{L,e}}{\lambda_{c,{\rm HI}}} = 
3.5\cdot 10^{-11} \frac{n_{\rm HI}}{n^{1/2}_{{\rm H},ICM}} \; .
\end{equation}     
 
For weaker B fields than in gas pressure equilibrium the ratio can 
become much larger, i.e. the free electron motion is less hampered. 
But this only holds for the fraction of electrons moving perpendicular to 
the B vector, while electrons moving parallel to the B field are unaffected. 
\cite{mk01}, however, showed that random magnetic fields also reduce
the diffusivity of electrons traveling along the B field lines. This reduction 
of free electron motion and thus heat conduction by magnetic fields can 
also be parametrized by $\Phi_s$ as in eq.\ref{qsat}.

\begin{appendix}

\section{Axis effects (model U$^\star$)}

The numerical models in this paper are calculated assuming cylindrical
symmetry. This method has numerical artefacts near
the symmetry axis. Material, once fallen onto this axis, cannot be removed
from it because of mirrored conditions, i.e. zero gradients.
As a result a nozzle is formed at the front of
the cloud, which Dinge (\cite{d97}) interprets as RT instability.
On the other hand, ablated material that is pushed towards the axis
by vortices behind the cloud builds up elongated spurs. In contrast,
in 3D simulations these clumps of a typical thickness of a few cells oscillate
in the slipstream of the cloud because of asymmetrical instabilities.
In order to test the influence of these artefacts the evolution of
the clouds we switched to a 2D Cartesian grid and so removed the
symmetry axis. Fully 3D simulations with sufficient spatial resolution
including heat conduction are still too time consuming to be produced.
The initial model of the simulated clouds are therefore no longer spheres 
but infinite cylinders. As a consequence, the velocity field, the gravitational
potential and the density distribution of the initial model differ from the
axisymmetric case but are close to model U.

      \begin{figure*}[ht]
      \psfig{figure=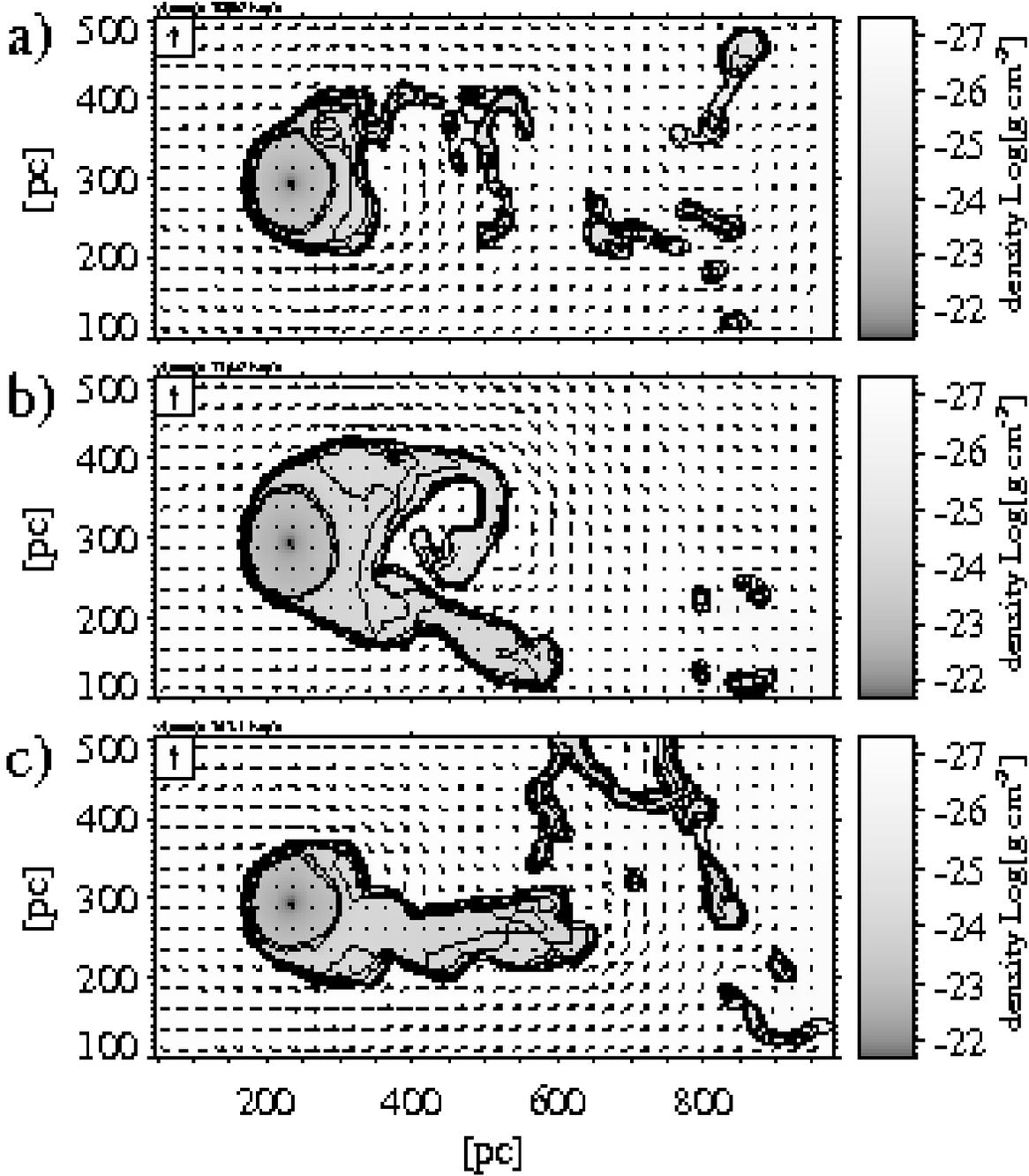,height=18.0cm,width=13.0cm,bbllx=14pt,bblly=14pt,bburx=404pt,bbury=509pt}
      \caption{\label{f14}Evolution of the density distribution for the 2d
      simulation using a cartesian grid with heat conduction.
      The evolution is shown at the times 25 Myr (upper
      panel), 50 Myr (middle) and 75 Myr (lower). Arrows indicate gas
      velocities scaling linearly with respect to the maximum velocity shown in the
      upper left. The contour lines represent 5, 10, 50, 100,\ldots $\times
      \rho_{\mbox{\tiny ISM}}$. The thick contour encloses the gravitationally
      bound part of the cloud.}
      \end{figure*}

The evolution of this model U$^\star$ is shown 25 Myr, 50 Myr and
75 Myr after the beginning of the calculations with consideration of
heat conduction (Fig.~\ref{f14}). Because of the
infinite cylinder instead a sphere, the velocity of the stream
near the top and the bottom of the cloud is much higher than in the
simulations shown before. As a result, the onset of KH instability is
facilitated. The tail of ablated cloud material at the rear
of the cloud can wave in the stream. Regions of the tail that become
too dilute are cut. Nevertheless a massive tail similar to model U
forms. No spur forms at the front of the cloud and must
therefore be identified as an artefact of the axisymmetric conditions.

Although there are some differences in the appearance of the
clouds in cylindrical and Cartesian geometry, the global formation of 
a head-tail structure of the cloud remains the same in both geometries. 
The thin elongated structures stripped off the cloud 
in the Cartesian grid are comparable to the material that get stuck on
the symmetry axis in cylindrical symmetry. In both cases the filaments
leave the computational domain and do not influence the evolution of the 
gravitationally bound cloud. Therefore the use of a cylindrical symmetry is 
justified.

      \section{Stability analysis}

      \subsection{Rayleigh-Taylor instability}

If a cool, dense cloud moves through a hot, tenuous gas, its surface
will become subject to KH and RT instabilities.
If the cloud is massive enough so that the effective
gravity is directed toward the cloud center, RT instabilities can be
suppressed. The growth rate for the RT instability in the absence of
a gravitational field is expressed as
(Chandrasekhar \cite{c61})

      \begin{equation}
      |\omega^2|_{\mbox{\tiny RT}} =
      a k \frac{\rho_{\mbox{\tiny cl}}-\rho_{\mbox{\tiny ISM}}}
      {\rho_{\mbox{\tiny cl}}+\rho_{\mbox{\tiny ISM}}} \quad ,
      \end{equation}

where $k$ is the wavenumber of the perturbation, $a$ the acceleration of the
cloud due to the wind, $\rho_{\mbox{\tiny cl}}$ and $\rho_{\mbox{\tiny ISM}}$
are the densities of the cloud and of the streaming medium,
respectively. This means that
perturbations with shorter wavelengths grow faster than larger ones. As long
as the wavelength is small compared to the cloud radius, the net effect
of RT instability is a small-scale mixing of the ISM with the cloud but does not
lead to its disruption. The cloud is stabilized against RT by a gravitational
field and RT is totally suppressed if the gravitational acceleration $g$ is
larger than the acceleration due to the wind (Murray et al. \cite{m93}).

      \subsection{Kelvin-Helmholtz instability}

In order to study the influence of KH instability we perturb the 
incompressible fluid equations

      \begin{equation}
      \rho \frac{\partial v_i}{\partial t} + \rho
      \sum_{j=1}^{3} v_j \frac{\partial v_i}{\partial x_j} =
      - \rho g_i - \frac{\partial p}{\partial x_i} \qquad \quad [i=1,2,3]
      \end{equation}
      \begin{equation}
      \frac{\partial \rho}{\partial t} + \sum_{i=1}^{3} v_i
      \frac{\partial \rho}{\partial x_i} = 0
      \end{equation}
      and
      \begin{equation}
      \sum_{i=1}^{3} \frac{\partial v_i}{\partial x_i} = 0
      \end{equation}

for a solenoidal velocity field according to Chandrasekhar~(1961).
The quantities density $\rho$, pressure $p$ and velocity $\vec{v}$, are
replaced by the perturbed ones $\rho + \delta\rho$, $p + \delta p$ and
$\vec{v} = \vec{v_0}+\delta\vec{v}$. The acceleration due to the gravitational
force $\vec{g}$ remains unperturbed. The initial velocity field $\vec{v_0}$
for the KH stability analysis is zero except for $v_x(z)$ which is the flow
velocity and therefore an arbitrary function of the height $z$.
After linearization we have five partial differential equations (PDEs) for
the five perturbed quantities. We now
consider perturbations of the form

      \begin{equation}
      \delta X \propto \exp[ i ( k_xx + k_yy + \omega t) ] \; .
      \end{equation}

Inserting this into the five PDEs and combining them we have a differential
equation for the z-dependence of the perturbation $\delta v_z$ which is
perpendicular to the streaming $v_x$

      \begin{eqnarray}{\label{gl1}}
      \lefteqn{
      \frac{\mbox{d}}{\mbox{dz}} \left[ \rho ( \omega + k_x v_x )
      \left( \frac{\mbox{d} \delta v_z}{\mbox{dz}} \right)
      -\rho k_x \, \delta v_z \left( \frac{\mbox{d} v_x}{\mbox{dz}} \right) \right] }
      \nonumber \\
      & &- k^2 \rho \: \delta v_z ( \omega + k_x v_x )
      =
      g k^2 \left( \frac{\mbox{d} \rho}{\mbox{dz}} \right)
      \frac{\delta v_z}{ \omega + k_x v_x }  \;   .
      \end{eqnarray}

This equation can be solved analytically
for two homogeneous fluids of densities $\rho_{\mbox{\tiny ISM}}$ and
$\rho_{\mbox{\tiny cl}}$ separated by a horizontal boundary and
moving with horizontal velocities $v_{\mbox{\tiny ISM}}$
and $v_{\mbox{\tiny cl}}$. Starting with a stable
configuration, i.e. the density of the
upper fluid $\rho_{\mbox{\tiny ISM}}$ is less than the lower one, 
we obtain the dispersion relation for $\omega(k)$

      \begin{equation}
      \omega(k) =
      \sqrt{\frac{ k^2 \rho_{\mbox{\tiny cl}} \rho_{\mbox{\tiny ISM}}
      (v_{\mbox{\tiny cl}} - v_{\mbox{\tiny ISM}})^2 -
      g k (\rho_{\mbox{\tiny cl}}^2 - \rho_{\mbox{\tiny ISM}}^2) }
      {(\rho_{\mbox{\tiny cl}} + \rho_{\mbox{\tiny ISM}})^2} }  \;   .
      \end{equation}

Instability occurs for real values of $\omega(k)$, so there is a minimal
wavenumber $k_{\mbox{\tiny min}}$ above which no stable mode exists.
With $v_{\mbox{\tiny rel}} = v_{\mbox{\tiny cl}}-
v_{\mbox{\tiny ISM}}$ and $\rho_{\mbox{\tiny cl}} \gg
\rho_{\mbox{\tiny ISM}}$ this upper limit for stability reads:

      \begin{equation}
      k_{\mbox{\tiny min}} = \frac{g \rho_{\mbox{\tiny cl}} }
      {\rho_{\mbox{\tiny ISM}} \: v_{\mbox{\tiny rel}}^2} \; .
      \end{equation}

For unstable wavenumbers the timescale for the exponential increase of
the perturbation reads

      \begin{equation}{\label{kh}}
      \tau_{\mbox{\tiny KH}} = \frac{1}{\omega(k)} \;   .
      \end{equation}

Because most destructive perturbations are those with wavelengths
of the order of the cloud radius ($\lambda = R_{\mbox{\tiny cl}} =
2 \pi / k$) there is a critical gravitational acceleration
$g_{\mbox{\tiny crit}}$ above which
the cloud is stabilized:

      \begin{equation}
      g_{\mbox{\tiny crit}}=\frac{2 \pi v^2_{\mbox{\tiny rel}}
      \rho_{\mbox{\tiny ISM}}}
      {\rho_{\mbox{\tiny cl}} R_{\mbox{\tiny cl}}} \;  .
      \end{equation}

If a spherical cloud with uniform
density is assumed, its gravitational acceleration is

      \begin{equation}
      g=\frac{G M_{\mbox{\tiny cl}}}
      {R^2_{\mbox{\tiny cl}}}
      \qquad \mbox{and} \qquad
      M_{\mbox{\tiny cl}} = 4/3 \,\pi \rho_{\mbox{\tiny cl}}
      R^3_{\mbox{\tiny cl}} \quad .
      \end{equation}

This leads to a critical mass of the cloud

      \begin{equation}{\label{mlow}}
      M_{\mbox{\tiny lower}} = \sqrt{6} \pi
      \frac
      {v^3_{\mbox{\tiny rel}}}
      {\sqrt{\rho_{\mbox{\tiny cl}}}}
      \left (
      \frac
      {\rho_{\mbox{\tiny ISM}}}
      {G \rho_{\mbox{\tiny cl}}}
      \right )^{3/2}
      \end{equation}

above which KH instabilities are suppressed. Clouds with
$M_{\mbox{\tiny lower}}$ $< M_{\mbox{\tiny cl}}$ $< M_{\mbox{\tiny max}}$
should survive long enough so that
star formation can occur.
These estimates are valid for the case of
density and velocity jumps at the cloud boundary.

\begin{figure}[ht]
      \psfig{figure=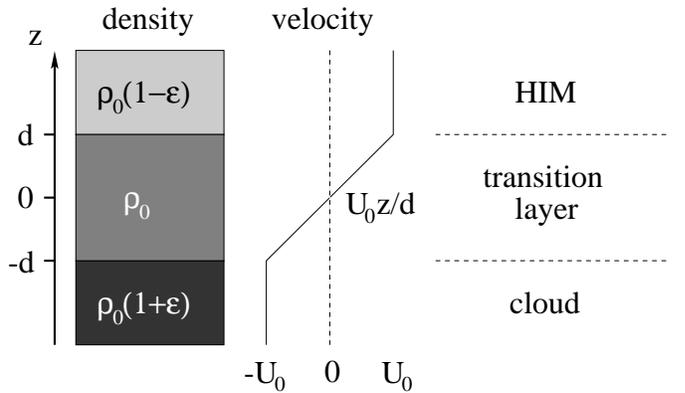,width=8.8cm,angle=-90}
      \caption{\label{sb1}Density and velocity structure for the
      stability study: a cloud moving with $2U_0$ relatively to a hot
      gas with density contrast $(1+\epsilon)/(1-\epsilon)$.}
      \end{figure}

If we take dissipative forces like heat conduction or viscosity into account,
which are able to diminish the density and velocity gradient, the picture
changes drastically. For further investigations we introduce an
artificial transition
layer of thickness $2d$ and density $\rho_0$ in which the
velocity decreases linearly from the value $U_0$ outside the cloud ($z>d$)
to $-U_0$ inside the cloud ($z<-d$) (see Fig.~\ref{sb1}). The ISM at
velocity $U_0$ has a density $\rho_0 \cdot (1 - \epsilon)$, the cloud
material with $-U_0$ has $\rho_0 \cdot (1 + \epsilon)$.

For a stability analysis we solve Eq.~\ref{gl1} together with the
boundary conditions
at the transition zone. The relation between $\omega$ and $k$ as a
function of $d$ and $\epsilon$ is given by

      \begin{eqnarray}{\label{gl2}}
      \lefteqn{
      \exp[-2 \, \kappa] = \left[ 1 - \frac{\kappa (\nu +1)^2}
      {J + (\nu + 1) + 0.5 \, \epsilon \, \kappa (\nu+1)^2} \right] }
      \nonumber \\
      & &\times \left[ 1 - \frac{\kappa (\nu -1)^2}
      {J - (\nu - 1) - 0.5 \, \epsilon \, \kappa (\nu-1)^2} \right]
      \end{eqnarray}
      with the Richardson number $J$ for this problem
      \begin{equation}
      J = \frac{\epsilon g d}{U_0^2} , \qquad \nu = \frac{\omega}{k U_0} \qquad
      \mbox{ and } \qquad \kappa = 2kd  \;  .
      \end{equation}

Modes with wavenumbers $k$ are stable if there are
no complex solutions for $\omega$. The case $\epsilon \rightarrow 0$ is
discussed in Chandrasekhar (\cite{c61}). For arbitrary $\epsilon$
we solved Eq.~(\ref{gl2}) numerically, yielding an instability strip
in the $k$-space. For some combinations of $d$ and $\epsilon$,
most destructive perturbations with wavelengths
of the order of the cloud radius can be stabilized. 

\begin{figure}[ht]
      \psfig{figure=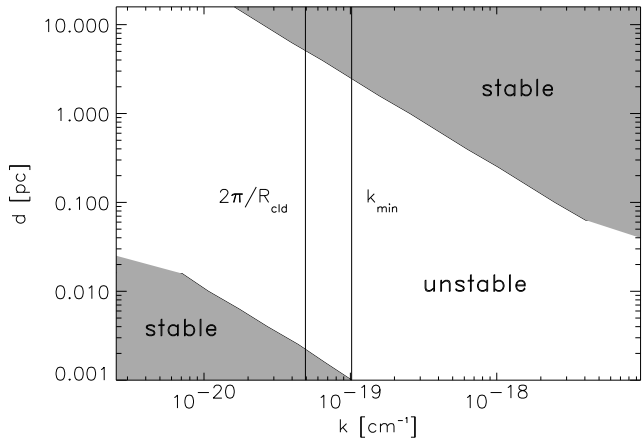,height=6.1cm,width=8.8cm,bbllx=20pt,bblly=345pt,bburx=555pt,bbury=715pt}
      \caption{\label{sb2} Kelvin-Helmholtz instability strip for 
      Model U with a density contrast $\epsilon$ of $1.2 \cdot 10^4$ 
      at various wave numbers $k$ and extents of the transition layer d.}
      \end{figure}

As an example, we investigate a model with a density 
contrast of $1.2 \cdot 10^4$, comparable to the Model U in subsequent
calculations, which corresponds to $\epsilon$ = 0.99983. For different
values of the transition zone thickness $d$, the consequence of 
perturbations by certain wave vectors $k$ is calculated and shown in
Figure~\ref{sb2}. 
From this, two facts become obvious: clouds that 
are unstable for certain $k$ are stabilized if the transition zone thickens. 
In this particular case, e.g., perturbations with wavelengths 
on the order of the cloud radius ($k = 2 \pi / R_{cl}$) that are usually
most destructive, are damped for transition layer thicknesses $d$ of more 
than 4 pc (upper right zone in Fig.~\ref{sb2}). 
Secondly, self-gravity leads to smaller $d$s and, by this, to stability 
for low $k$s (lower left region in Fig.~\ref{sb2}). For $k = 2 \pi / R_{cl}$ 
this model is stable for $d$ smaller than 0.002 pc. 

Spatially poorly resolved numerical simulations produce artificially 
thick boundary layers due to numerical noise. Since such extensions 
would put the model into the instability strip, KH instability sets in. 
Heat conduction broadens the boundary layer to values $d >$ 4 pc 
so that the cloud becomes stabilized. 

\end{appendix}

      \begin{acknowledgements}
      The authors thank Tim Freyer, Miguel Avillez, and Dieter Breitschwerdt
      for stimulating discussions, 
      and Tomek Plewa for providing us with his numerical code solving the
      heat conduction equation. The authors are grateful for very constructive
      comments of an anonymous referee. This work was partly supported by the
      Deutsche Forschungsgemeinschaft (DFG) under grant numbers HE~1487/5-3
      and HE~1487/25-1. The computations were performed at the Rechenzentrum
      der Universit\"at Kiel, the Konrad-Zuse-Zentrum f\"ur Informationstechnik
      in Berlin, and the John von Neumann-Institut f\"ur Computing in
      J\"ulich.
      \end{acknowledgements}


      \end{document}